\documentclass[aps,prl,twocolumn,superscriptaddress,groupedaddress]{revtex4}
\usepackage[english]{babel}
\usepackage{epstopdf}
\usepackage{pdfpages}
\usepackage{float}
\usepackage{enumerate}

\usepackage{amsmath} 
\usepackage{dsfont}
\usepackage{cancel}
\usepackage{amsthm} 
\usepackage{amssymb}	
\usepackage{empheq}
\usepackage[absolute,overlay]{textpos}
\usepackage{xcolor}

\allowdisplaybreaks

\let\vaccent=\v 
\renewcommand{\v}[1]{\ensuremath{\mathbf{#1}}} 
 
\let\baraccent=\= 
\renewcommand{\=}[1]{\stackrel{#1}{=}} 
\DeclareMathOperator{\Tr}{Tr}

\newcommand{\beginsupplement}{%
        \setcounter{table}{0}
        \renewcommand{\thetable}{S\arabic{table}}%
        \setcounter{figure}{0}
        \renewcommand{\thefigure}{S\arabic{figure}}%
     }

\begin{document}

\title{Generating quantum multi-criticality in topological insulators by periodic driving}
\date{\today}

\author{Paolo Molignini}
\affiliation{Institute for Theoretical Physics, ETH Z\"{u}rich, 8093 Zurich, Switzerland}
\author{Wei Chen}
\affiliation{Institute for Theoretical Physics, ETH Z\"{u}rich, 8093 Zurich, Switzerland}
\affiliation{Department of Physics, PUC-Rio, 22451-900 Rio de Janeiro, Brazil}
\author{R. Chitra}
\affiliation{Institute for Theoretical Physics, ETH Z\"{u}rich, 8093 Zurich, Switzerland}

\begin{abstract}
We demonstrate that the prototypical two-dimensional Chern insulator hosts exotic quantum multi-criticality in the presence of an appropriate periodic driving: a linear Dirac-like transition coexists with a nodal loop-like transition caused by emerging symmetries. The existence of multiple universality classes and scaling laws can be unambiguously captured by a single renormalization group approach based on the stroboscopic Floquet Hamiltonian, regardless  of whether the topological transition is associated with the anomalous edge modes or not.
\end{abstract}

\maketitle

\textit{Introduction} --- Periodic driving provides an unprecedented channel to engineer exciting nonequilibrium quantum phases. Of particular interest are the various topologically ordered phases that can be achieved by periodic driving, such as Floquet topological insulators~\citep{Kitagawa, Lindner, Cayssol:2013, Harper:2017, Roy:2017,Esin:2018}, Floquet topological superconductors~\cite{Liu,Thakurathi, Thakurathi:2014, Wang:2014,Sacramento:2015,Thakurathi:2017,Molignini:2017,Molignini:2018,Cadez:2019}, and various exotic Floquet semimetals~\cite{Gong:2016, Gong:2016-2, Gong:2016-3, Bucciantini:2017, Huebener:2017, Cao:2017, ShuChen:2017, ChenZhou:2018}, as well as nodal lines or loops~\cite{Li:2018}. 
Most importantly, this plethora of Floquet topological phases can be very efficiently tuned by simple manipulations of the drive. This versatility provides an unrivalled opportunity to investigate the quantum criticality near the topological phase transitions (TPTs).

In this Letter, we demonstrate the feasibility of inducing TPTs belonging to multiple universality classes, as well as quantum multi-criticality, by means of periodic driving in an otherwise ordinary, noninteracting topological insulator. 
This result is remarkable, given that in static systems with local Landau order parameters these features  usually arise from the complicated interplay between various interacting and kinetic energy scales~\cite{multicriticality-book,Zacharias:2009,Carr-book,Pixley:2014, Brando:2016}.
The identification of this unexpected multi-criticality is based on a unified framework that views the topological order as an momentum space integration of a curvature function~\cite{Chen:2016,Chen-Sigrist:2016,Kourtis:2017,Chen:2018}, and the generic feature that the curvature function and edge state decay length diverge~\cite{Rufo:2019} at the TPTs, from which the existence as well as  coexistence of multiple universality classes and the scaling laws are uncovered. 

To demonstrate the aforementioned  features, we employ a prototypical two-dimensional (2D) Floquet-Chern insulator (FCI) that exhibits additional peculiar features that are themselves of extreme interest. 
Firstly, this system hosts so-called anomalous phases --- with no static counterpart --- that break the ubiquitous bulk-edge correspondence: they are characterized by a trivial stroboscopic bulk topology, while still exhibiting anomalous edge modes (AEMs) in the quasienergy spectrum~\cite{Rudner:2013, Mukherjee:2017, Yao:2017}. 
We establish that anomalous TPTs can be well captured by stroboscopic physics and do not require knowledge of the full-time dependence (micromotion). Secondly, we uncover that an appropriate driving induces a \textit{nodal loop semimetal} (NLS) due to an emergent chiral mirror symmetry, which corresponds to elusive topological nodal loop band inversions studied in the context of the spin Hall effect~\cite{Qi:2006,Li:2016}. 
This suggests the feasability of realizing exotic symmetry-induced topological states by choosing specific driving strategies.
We will demonstrate that the unified scheme based on the momentum space curvature function allows the application of a curvature renormalization group (CRG) approach~\citep{Chen:2016,Chen-Sigrist:2016}, which has been successfully applied to determine the phase boundaries in numerous interacting and noninteracting models, both static and periodically driven~\citep{Chen:2016,Chen-Sigrist:2016,Kourtis:2017,Chen:2018, Molignini:2018}. The CRG approach based on the stroboscopic Floquet Hamiltonian unambiguously captures the TPTs despite all the richness of multi-criticality, AEMs, and emergent nodal loops. 

\begin{figure}
\centering
\includegraphics[width=\columnwidth]{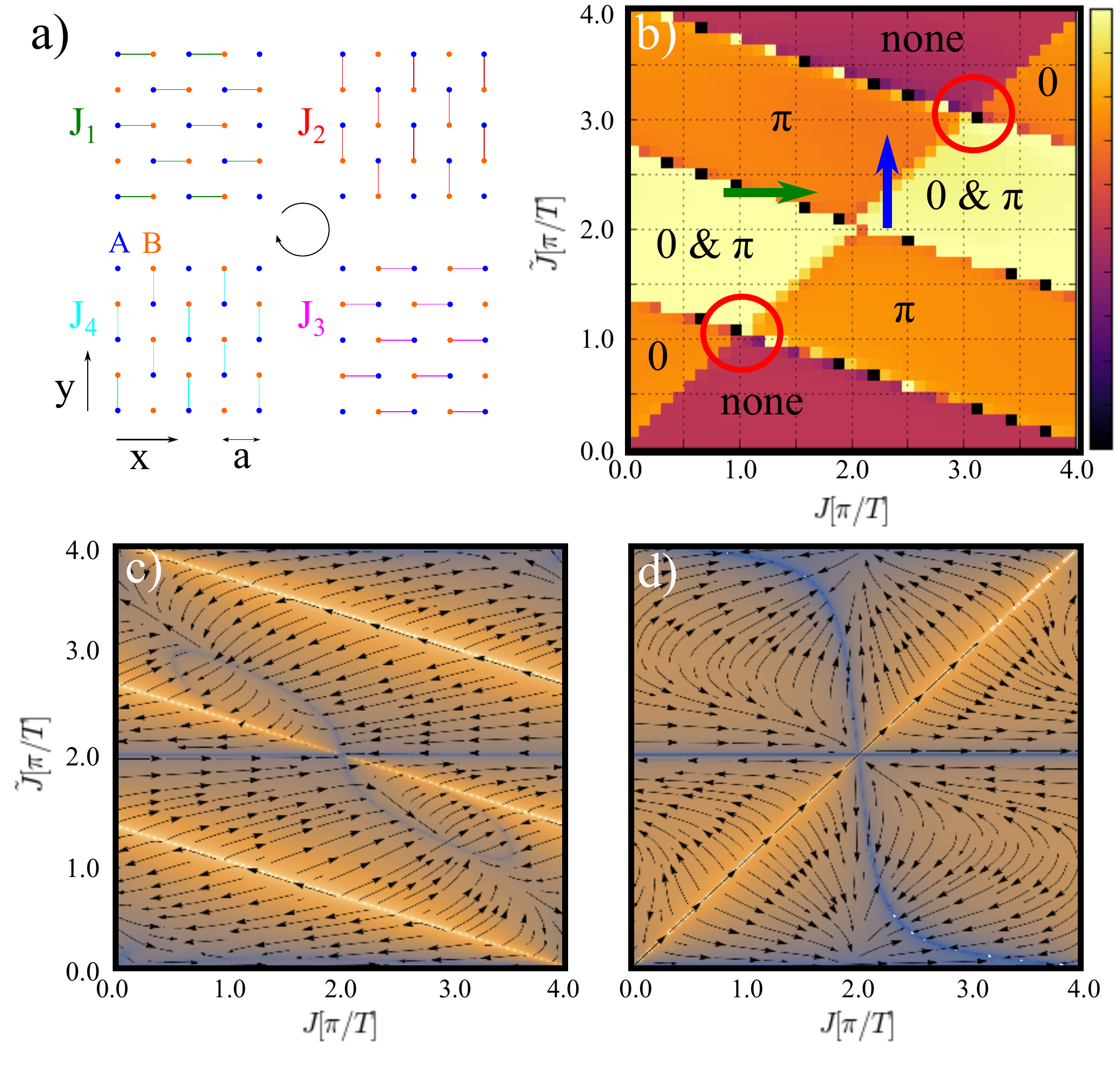}
\caption{a) Depiction of the periodically-driven 2D tight-binding model discussed in the main text. 
We consider a periodic four-step driving protocol for which the first hopping is different  from the next three, $J \equiv J_1 \neq J_2=J_3=J_4 \equiv \tilde{J}$.
b) Phase diagram obtained from the time-integrated topological invariant, indicating the type of topological excitations in each phase.
The arrows, also shown in Fig.~\ref{fig:curv-func-2D}, indicate the two classes of TPTs studied, while the red circles highlight multi-critical points.
c)-d) CRG flow diagram evaluated at $\mathbf{k}_0=(0,0)$ and $\mathbf{k}_0=(0,\pi)$ along direction $k_i=k_x$. 
The color codes indicate the log of the numerator of the CRG equation, $\log \left[ \partial_{k_i}^2 F(\mathbf{k}_0, \mathbf{M}) \right]$, with yellow being high values (critical lines) and blue low values (fixed lines). 
Choosing other HSPs (\textit{i.e.} $(\pi,0)$ and $(\pi,\pi)$) or the direction $k_i=k_y$ leads to similar flow diagrams.}
\label{fig:unequal-hopping-PD-CRG}
\end{figure}

\textit{Model} --- We  consider a paradigmatic 2D FCI already realized in photonic lattices~\cite{Rudner:2013, Mukherjee:2017}.
The model describes fermions with \textit{modulated} nearest-neighbor hoppings on a square lattice:
\begin{align}
{\cal H}(t) &= \sum_{\mathbf{k}=(k_x, k_y)} \left( \begin{array}{cc} c_{\mathbf{k}, A}^{\dagger} & c_{\mathbf{k}, B}^{\dagger} \end{array} \right) H(\mathbf{k}, t) \left( \begin{array}{c} c_{\mathbf{k}, A} \\ c_{\mathbf{k}, B} \end{array} \right) \nonumber \\
H(\mathbf{k}, t) &= - \sum_{n=1}^4 J_n(t) \left( e^{i \mathbf{b}_n \cdot \mathbf{k}} \sigma^+ + e^{-i \mathbf{b}_n \cdot \mathbf{k}} \sigma^- \right).
\end{align}
The lattice vectors $\mathbf{b}_1 = -\mathbf{b}_3 = (a,0)$ and $\mathbf{b}_2=-\mathbf{b}_4 = (0,a)$ connect the sublattices $\alpha=A,B$ on which the creation or annihilation operators $c_{\mathbf{k},\alpha}^{(\dagger)}$ act, and $\sigma^{\pm}$ refer to Pauli matrices.
A schematic illustration of the model is provided in Fig.~\ref{fig:unequal-hopping-PD-CRG}.
The hoppings $J_n(t)$ are periodically modulated with a four-step protocol of period $T$, in which during the $n$-th step of the cycle only the hopping $J_n$ in direction $\mathbf{b}_n$ is active~\cite{Mukherjee:2017}. 
Furthermore, we consider the special case of  $J \equiv J_1 \neq J_2=J_3=J_4 \equiv \tilde{J}$ (see Fig.~\ref{fig:unequal-hopping-PD-CRG} for clarity).
Note that this type of modulation inherently introduces a circular pattern of hoppings that enables the propagation of edge modes in a strip geometry~\cite{Rudner:2013, Mukherjee:2017}.

In Ref.~\cite{Rudner:2013}, it was shown that stroboscopic dynamics is insufficient to determine the correct number of edge modes generated in this model.
Micromotion, \textit{i.e.} time evolution within a period, should be considered explicitly by introducing a computationally cumbersome time-integrated topological invariant determined from the full time evolution operator $U(t, 0) = \mathcal{T} \exp \left[ -i \int_0^t H(t') \mathrm{d} t' \right]$~\cite{Rudner:2013}\footnote{Throughout our work we have set $\hbar=1$}.
In general, micromotion plays a crucial in role in the definition and determination of the correct topological invariants~\cite{Unal:2019,Unal:2019-2}.
However, information about the TPTs is readily determined by the gap closures in the quasienergy spectrum at zero and $\pi$ quasienergies, which can be readily extracted from the stroboscopic bulk effective Hamiltonian $h_{\text{eff}}(\mathbf{k})$~\cite{Molignini:2018}.  
For a general unitary $2 \times 2$-Floquet operator $U_{\mathbf{k}}(T,0)= \left( \begin{array}{cc} A(\mathbf{k}) & B(\mathbf{k}) \\ -B(\mathbf{k})^* & A(\mathbf{k})^* \end{array} \right)$, the effective Floquet Hamiltonian defined via  $U_{\mathbf{k}}(T,0)= e^{-i h_{\text{eff}}(\mathbf{k}) T}$ takes the following form~\cite{Molignini:2018, supmat}:
\begin{equation}
h_{\text{eff}}(\mathbf{k}) \propto  {\rm {Im}}[B] \sigma^x + {\rm Re}[B] \sigma^y + {\rm Im}[A] \sigma^z  \equiv \mathbf{n}(\mathbf{k}) \cdot \boldsymbol{\sigma}.
\label{strobo-hamiltonian}
\end{equation}
The quasienergy dispersions of the Floquet bands rescaled to $[-\pi,\pi)$ are then obtained as the eigenvalues of the operator $h_{\text{eff}}(\mathbf{k}) T$.
The topology of the stroboscopic bulk effective Hamiltonian is  mapped by the Chern number~\cite{ChiuReview:2016} which for the two-band system analyzed here is $\mathcal{C} = - \frac{1}{2\pi} \int_{-\pi}^{\pi} \mathrm{d}k_x \int_{-\pi}^{\pi} \mathrm{d}k_y  \: F(\mathbf{k})$, where 
\begin{equation}
F(\mathbf{k}) = \hat{\mathbf{n}}(\mathbf{k}) \cdot \left[ \partial_{k_x} \hat{\mathbf{n}}(\mathbf{k}) \times \partial_{k_y} \hat{\mathbf{n}}(\mathbf{k}) \right]
\label{Berry-curvature-strobo}
\end{equation}
is the Berry curvature of the Floquet band in question.

\textit{Detection of TPTs} --- The Chern number $\mathcal{C}$ of the effective bulk Hamiltonian counts the difference between the numbers of edge modes above and below each Floquet band~\cite{Rudner:2013}.
If edge modes exist only at quasienergy $0$ \textit{or} $\pm \pi$,  $\mathcal{C}$ of each band will correctly capture the number of edge modes. 
However, for coexisting $0$ \textit{and} $\pm \pi$ modes  $\mathcal{C}=0$, the bulk-edge correspondence is broken, and we have a phase with AEMs~\cite{Rudner:2013}. 
The full phase diagram based on time-integrated topological invariants as defined in Ref.~\cite{Rudner:2013} is  mapped in Fig.~\ref{fig:unequal-hopping-PD-CRG}b) as a function of the two hopping strengths $J$ and $\tilde{J}$.
We can distinguish multiple phases which are either topologically trivial, hosting only one edge mode at quasienergy $0$ or $\pi$, or hosting AEMs. 
The topological phase boundaries have an analytical form given by $\tilde{J}=J$ and $\tilde{J} = \frac{1}{3} \left( n - J \right)$, where $n \in \left[1,2,3\right]$~\cite{Mukherjee:2017}.

Using the CRG method, we now show that information about the TPTs and their criticality can be extracted even in the presence of AEM from the stroboscopic Berry curvature alone, even though its integration $\mathcal{C}$ is not necessarily equal to the true time-integrated topological invariant.
We first briefly summarize the method developed originally for static systems in \citep{Chen:2016,Chen:2017,Chen-Sigrist:2016,Kourtis:2017,Chen:2018} and for Floquet systems in \cite{Molignini:2018}. Generally, at any system parameters ${\bf M}=(J, \tilde{J})$, the  curvature function peaks around high-symmetry points (HSPs) $\mathbf{k}_0$ satisfying $\mathbf{k}_0 = - \mathbf{k}_0$. Across the topological phase boundaries described by $\{ \mathbf{M}_c \}$, the peaks diverge and flip sign to preserve the quantization of the Chern number~\citep{Chen:2016, Chen-Sigrist:2016, Chen:2017}.
The CRG relies on the scaling procedure $F(\mathbf{k}_0, {\bf M}^{\prime}) = F(\mathbf{k}_0 + \delta \mathbf{k}, {\bf M})$ that searches the trajectory in the parameter space along which the diverging peak $F(\mathbf{k}_{0},{\bf M})$ is gradually flattened~\citep{Chen:2016, Chen-Sigrist:2016, Chen:2017}. Defining $\mathrm{d} l \equiv \delta k^2$, $dM_{i}=M_{i}'-M_{i}$, the scaling procedure yields the RG equation
\begin{equation}
\frac{\mathrm{d}M_{i}}{\mathrm{d} l} = \frac{1}{2} \frac{\partial^2_{k_j} F(\mathbf{k}, {\bf M}) \big|_{\mathbf{k}=\mathbf{k}_0}}{\partial_{M_{i}} F(\mathbf{k}_{0}, {\bf M})}.
\label{generic_RG_equation}
\end{equation}
The topological phase diagram can be easily ascertained by analyzing the critical points of \eqref{generic_RG_equation}~\cite{supmat}.
Furthermore, the criticality of the TPT is characterized by the divergences of both $F(\mathbf{k}_0,\mathbf{M})$ and the concomitant inverse of the full width at half maximum in directions $i=x,y$ expressed as $\textit{FWHM}_i \equiv \frac{2}{\xi_{k_{0,i}}}$, see also Ref.~\citep{Chen:2016, Chen-Sigrist:2016, Chen:2017}.
As ${\bf M} \to {\bf M}_{c}$, these quantities diverge like $ F(\mathbf{k}_{0},{\bf M}) \propto |{\bf M}-{\bf M}_{c}|^{-\gamma}$ and $\xi_{\mathbf{k}_{0,i}}\propto |{\bf M}-{\bf M}_{c}|^{-\nu_i}$.
The conservation of $\mathcal{C}$ in a phase leads to precise scaling laws relating $\gamma$ and $\nu_i$.  For example, $\sum_i \nu_i =  1 + 1 = \gamma = 2$ for an isotropic 2D Dirac model~\citep{Chen:2017}.
The Fourier transform of the curvature function yields a {\it{Wannier state correlation function}}, which decays exponentially in real space due to the peak shape of the curvature function in momentum space, with $\xi_{\mathbf{k}_{0,i}}$ playing the role of the correlation length, and $F(\mathbf{k}_{0},{\bf M})$ serving as the analog of the susceptibility in the Landau paradigm~\cite{supmat}. The critical exponents $\left\{\gamma,\nu_i\right\}$ then characterize the universality class of the TPT.

\textit{Quantum criticality of the TPTs} ---
We now analyze the criticality of the different TPTs that exist in this model, including transitions to phases with AEMs. 
We apply the CRG method to the stroboscopic curvature function \eqref{Berry-curvature-strobo} at the representative HSPs $\mathbf{k}_0=(0,0),(0,\pi)$~\footnote{Because of the $C_4$ symmetry of the lattice, all other HSPs (\textit{e.g.} $\mathbf{k}_0=(\pi,\pi), (\pi,0)$ etc.) are obtained by a shift with the reciprocal vectors $\mathbf{b}_1=(\pi,\pi)$ and/or $\mathbf{b}_2=(-\pi,\pi)$ and yield the same results as the two different representative HSPs $\mathbf{k}_0=(0,0),(0,\pi)$.} to extract the criticality of the topological phase boundaries.
Firstly, via a straightforward  analysis of the CRG equations, all phase boundaries, including those delineating AEM phases are correctly captured by the RG flow (see Figs.~\ref{fig:unequal-hopping-PD-CRG}a-b).
Note that this approach requires very little computational effort compared to the evaluation of the time-integrated topological invariant, since solving Eq.~(\ref{generic_RG_equation}) only requires to calculate $F(\mathbf{k},{\bf M})$ at few momentum points.
The fixed lines, on the other hand, illustrate the regions where the correlation length is shortest, indicating relatively localized Wannier states.
In the following, we focus exclusively on the critical lines, and detail the two dramatically different critical behaviors uncovered in this system.
  
\begin{figure}[t]
\centering
\includegraphics[width=\columnwidth]{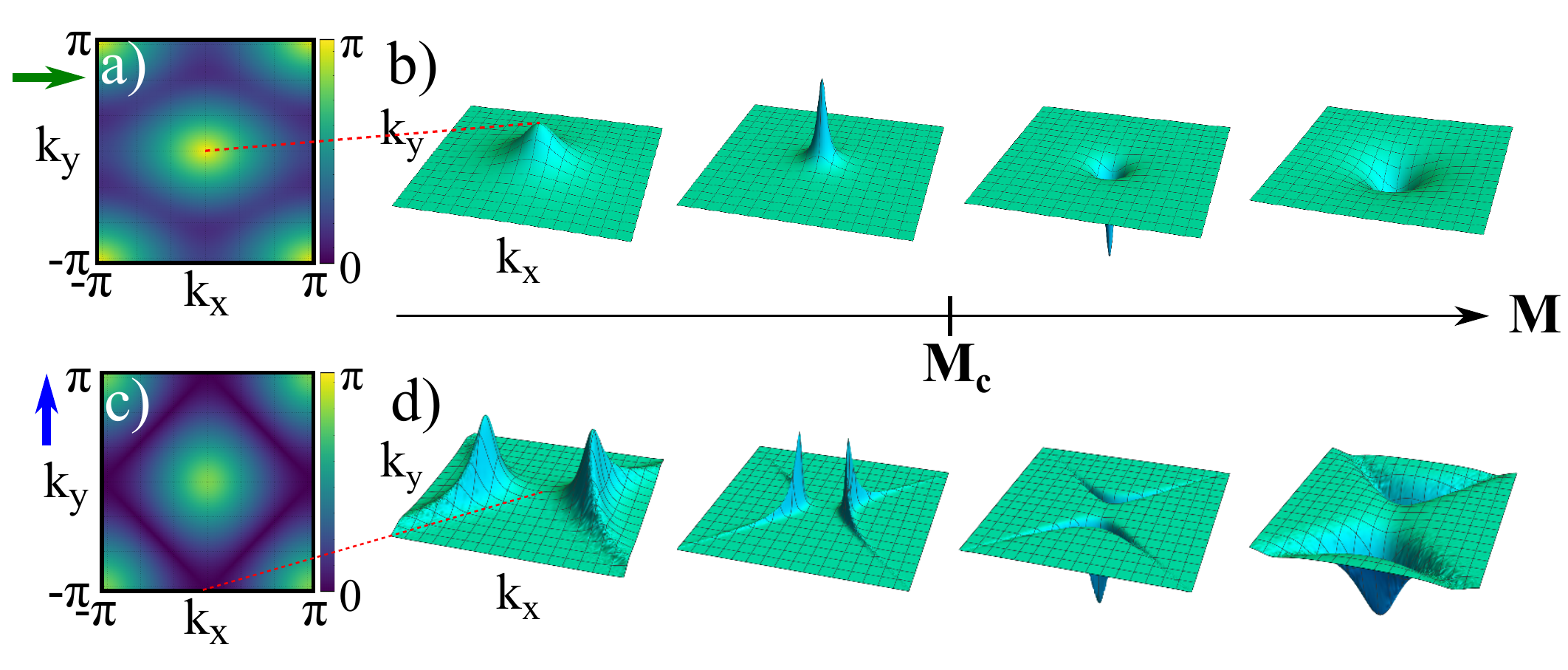}
\caption{Illustration of different TPTs (indicated by the arrows in Fig.~\ref{fig:unequal-hopping-PD-CRG}) in the FCI as a function of tuning parameter $\mathbf{M}=J$ (b) or $\mathbf{M}=\tilde{J}$ (d). a) Quasienergy dispersion exhibiting Dirac-cones with linear gap closure at quasienergy $\pi$ for $\tilde{J} = \frac{1}{3} \left(n - J \right)$. b) Behavior of the Berry curvature across the TPT with linear gap closure (Lorentzians).
c)  quasienergy dispersion with quadratic gap closure at quasienergy 0 for $\tilde{J} = J$~\cite{supmat}. 
d) Behavior of the Berry curvature across the TPT with quadratic gap closures, where non-Lorentzian \textit{pairs} of peaks flip sign \textit{and change direction}.}
\label{fig:curv-func-2D}
\end{figure}

\textit{Floquet-Dirac criticality} ---  Close to the $\tilde{J_c} = \frac{1}{3} \left( n - J \right)$ (\textit{e.g.}  green arrow in Fig.~\ref{fig:unequal-hopping-PD-CRG}b)) transitions, a Dirac-like linear gap closure at the Floquet band edge takes place at one of the HSPs $(k_x, k_y)=(0,0)$ and $(k_x, k_y)=(\pm \pi, \pm \pi)$, and the curvature function near the HSP has the shape of a single Lorentzian peak. 
As $\tilde{J} \to \tilde{J}_c$, the Lorentzian peak diverges and flips sign across the transition, with
critical exponents $\nu_{x}=\nu_{y}=1$, $\gamma=2$, fulfilling the scaling law $2 = \gamma = \nu_x + \nu_y$ (up to numerical accuracy).
This result implies that these TPTs belong to the same universality class of a static 2D isotropic Dirac model~\cite{Chen-Schnyder:2019}, and this critical behavior is independent of whether in the underlying phase AEMs exist or not.

\textit{Floquet nodal loop criticality} --- We now analyze the TPTs across $\tilde{J}_{c} = J$ (\textit{e.g.} blue arrow in Fig.~\ref{fig:unequal-hopping-PD-CRG}b)), for which the gap closure is revealed to be quadratic  along the nodal loops $k_y = \pm \pi \mp |k_x|$, realizing an elusive 2D  NLS.
For this nodal loop case, the curvature function is characterized by a pair of \textit{non-Lorentzian} peaks symmetrically shifted 
away from the HSP in $k_x$ or $k_y$ direction (depending on the direction of the transition), as shown in Fig.~\ref{fig:curv-func-2D}b). 
Across the TPT, the peaks simultaneously approach the HSP while diverging, then flip sign, and then depart again from the HSP \textit{but in the orthogonal direction}.
For simplicity, we fix the direction of the transition such that the peaks are along the $k_y$ direction before diverging.
We note that, because of the boomerang shape of the Berry curvature along the $k_x$-direction, the correlation length that correctly captures the conservation of the Chern number has to also be defined along the same curve, \textit{i.e.} as $\xi_{\mathbf{k}_{0,\tilde{x}}} \equiv \frac{2}{FWHM(\max_{k_y}\{ F(\mathbf{k})\})}$.
A fit of the curvature function for this geometry reveals $\gamma  =   \frac{3}{2}$,  $\nu_{\tilde{x}} = \frac{1}{2}$, and $\nu_{y} = 1$ (up to numerical accuracy).
These critical exponents fulfill the scaling law $\nu_{\tilde{x}} + \nu_{y} = \gamma$, but are distinct from Dirac models of any order of band crossing~\cite{Chen-Schnyder:2019}.

This different behavior clearly indicates that the TPT at $\tilde{J}_{c} = J$ belongs to a different universality class than the $\tilde{J_c} = \frac{1}{3} \left( n - J \right)$ transitions described earlier.
This is remarkable, as it is not customary for a single system to host two kinds of TPTs belonging to different universality classes.
Quite surprising is also that the two TPTs, described by drastically different effective theories, connect the same phases along two different parameter paths.

\begin{figure}[t]
\centering
\includegraphics[width=\columnwidth]{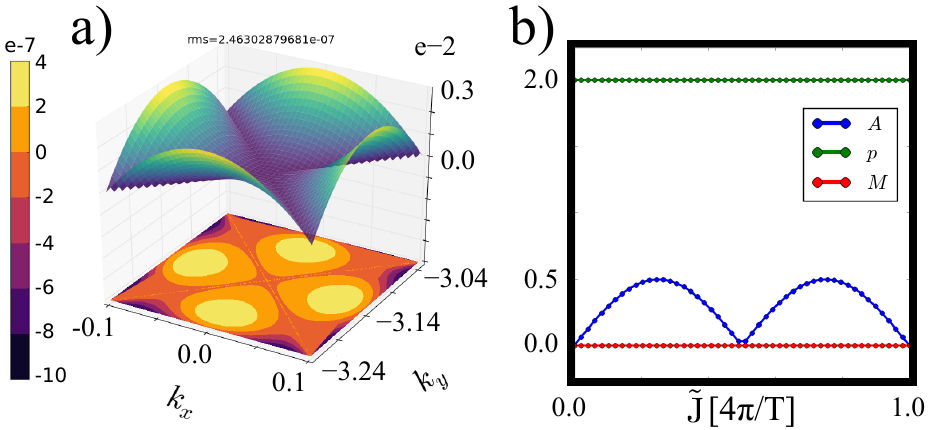}
\caption{a) Two-dimensional parametric fit of the quasienergy dispersion (upper band) to Eq. \eqref{energy-disp-nodal-loop} around the HSP $\mathbf{k}_0=(0,-\pi)$ for the TPT at $\tilde{J}=J=0.6$ in the FCI. The color bar on the left indicates the absolute error of the fit shown as a contour plot.
b) Dependence of the fitting parameters $A$, $p$, and $M$ on the position of the TPT $\tilde{J}=J$, showing that only the overall scaling $A$ changes along the transition line.
}
\label{fig:nodal-loop-mapping}
\end{figure}

We now present an effective theory for this  Floquet-engineered NLS  which  perfectly captures  this physics.
First note  that  in the vicinity of  $\tilde{J}_{c} = J$ transition line, the  quasienergy dispersion   around the HSPs $\mathbf{k}_0=(0, \pm \pi), (\pm \pi, 0)$ is  well described by the following:
\begin{equation}
E = \pm A \sqrt{((k_x-k_{0,x})^2 - (k_y-k_{0,y})^2 )^p + M^2},
\label{energy-disp-nodal-loop}
\end{equation}
where the parameters $A$, $p$, and $M$ are determined numerically. 
A fit of the quasienergy dispersion to the form of Eq.~\eqref{energy-disp-nodal-loop} is depicted in Fig.~\ref{fig:nodal-loop-mapping}a). 
Close to $\tilde{J}_{c} = J$,  $p=2$ and at the transition the mass term $M=0$.
Remarkably, along the entirety of the transition line $\tilde{J}_{c} = J$, the dispersion exhibits the same shape with the same values of $p$ and $M$ (except at the multi-critical point where the dispersion is identically zero), and only its overall scaling factor $A$ varies  as illustrated  in Fig.~\ref{fig:nodal-loop-mapping}b). 
Note that $A$ has no influence on the topology because it does not affect the gap closures.

The peculiar non-Dirac quadratic gap closure of Eq.~\eqref{energy-disp-nodal-loop} can be naturally generated along the entire $\tilde{J}=J$ transition line from a \textit{single} $2\times2$ NLS Hamiltonian
\begin{align}
H_{NL} &= (\mu- 2\eta(\cos k_x + \cos k_y))^2 \sigma^x \nonumber \\
&\quad + \alpha (\sin k_x - \sin k_y)( \cos k_x + \cos k_y) \sigma^y \nonumber \\
& \quad + \beta (\sin k_x +\sin k_y)( \cos k_x + \cos k_y) \sigma^z,
\label{nodal-loop-Ham}
\end{align}
where $\sigma^i$ are the Pauli matrices, $\alpha$, $\beta$, $m$, $\mu$, and $\eta$ are parameters.
For $\mu=0$ (and arbitrary values of the other parameters) the energy dispersion of Eq. \eqref{nodal-loop-Ham} in the vicinity of the HSPs exactly recovers Eq.~\eqref{energy-disp-nodal-loop}, with the overall scale $A=\eta$ and $M=0$.
This model can further reproduce all the features observed in the FCI along the $\tilde{J}=J$ transition, including the shape of the Berry curvature across the TPT, the value of the critical exponents, and symmetries~\cite{supmat}.
Surprisingly, Floquet driving realizes an extension of  the model discussed in Refs.~\cite{Qi:2006,Li:2016}  in the context of the spin quantum Hall phases. 
Though 3D NLS have been abundantly discussed in the literature
~\cite{Mullen:2015,Rhim:2015,Lim:2017,Kopnin:2011,RonghanLi:2016,Chan:2016,Wang:2017,JianpengLiu:2017},  physical  realizations of 2D NLS (proposed as excellent candidates for spintronics) remain elusive ~\cite{Burkov:2011,Yu:2017,Zhou:2019}.
Recently proposed candidate systems include interpenetrating kagome-honeycomb lattices~\cite{Lu:2017} and ferromagnetic monochalcogenide monolayers~\cite{Zhou:2019}. 
Here, we see that via a simple driving protocol on a square lattice, the resulting Floquet-engineered Hamiltonian perfectly realizes the full static model of a 2D NLS.

\textit{Emergent chiral mirror symmetry} ---  The existence of two different universality classes in a simple non-interacting model indicates that they must be related to different symmetries.
In fact, the effective nodal loop theory is characterized by an increased chiral mirror symmetry that can be traced back to the driving scheme. 
At $\tilde{J}=J$, the effective Floquet Hamiltonian (and correspondingly the low-energy theory Eq. \eqref{nodal-loop-Ham} with $\mu = 0$) fulfills
\begin{equation}
\mathcal{M} h_{\text{eff}}(k_x, k_y) \mathcal{M}^{-1} = - h_{\text{eff}}(k_y, k_x)
\end{equation}
with $\mathcal{M} = \sigma^y$.
This symmetry acts together with charge conjugation $\mathcal{C} h_{\text{eff}}(k_x, k_y) \mathcal{C}^{-1} = - h_{\text{eff}}(-k_x, -k_y)$, $\mathcal{C}= \sigma^z \circ \mathcal{K}$ ($\mathcal{K}$ represents complex conjugation), to topologically protect the nodal loop from mass terms in the $y$ and $z$ components~\cite{supmat}, while it is absent for the linear Dirac transitions along $\tilde{J_c} = \frac{1}{3} \left( n - J \right)$.
The chiral mirror symmetry has a direct correspondence in terms of a restricted form of the Floquet operator $U(\mathbf{k}, T) = U^{\mathbb{T}}(\mathbf{k}^{\mathcal{M}}, T)$, where the superscript $\mathbb{T}$ denotes the matrix transpose and $\mathbf{k}^{\mathcal{M}}$ are the $k$-coordinates transformed under the chiral mirror symmetry (in this case, $\mathcal{M}: k_x \leftrightarrow k_y$).
By Floquet engineering the stroboscopic time evolution to have such property, it is in principle possible to generate a whole hierarchy of effective Floquet Hamiltonians with chiral mirror symmetry that can stabilize nodal loop semimetal phases. 

\textit{Multi-critical points} --- Lastly, we consider  the three multi-critical points shown in the phase diagram.
Because the two kinds of TPTs of the model occur at different HSPs, the points at $\tilde{J}=J=\pi/T, 3\pi/T$ are found to exhibit both a peak divergence at $\mathbf{k}_0 = (0,0)$ and $\mathbf{k}_0 = (\pm \pi, \pm \pi)$ (linear gap closure at quasienergy $\pi$), and a double-peak divergence around $\mathbf{k}_0 = (0,\pm \pi)$ and $\mathbf{k}_0 = (\pm \pi,0)$ (quadratic gap closure at quasienergy $0$).
These multi-critical points therefore display a coexistence of both TPTs belonging to different universality classes. 
The points at $\tilde{J}=J=2\pi n/T$, $n \in [0,1,2]$ are instead not critical, because there the Floquet operator $U(\mathbf{k},T) = e^{i h_{\text{eff}}(\mathbf{k}) T}$ is precisely the identity and the quasienergy dispersion $h_{\text{eff}}(\mathbf{k}) T$ collapses to a flat band at 0. This  behavior at $\tilde{J}=J=2\pi/T$ is also detected  in the CRG flow, where fixed lines and critical lines meet. The loss of criticality around these points is further corroborated by the fit of the quasienergy dispersion (Fig.~\ref{fig:nodal-loop-mapping}), which highlights the flat band as $A \to 0$. 

 Finally, we remark that the multiple universality classes and the multi-criticality uncovered in the present work can be experimentally detected in cold atomic realizations of the model through quantum interference maps of the Berry curvature~\cite{Duca:2015} or force-induced wave-packet velocity measurements~\cite{Price:2012,Jotzu:2014}. 
Our work thus paves the path for future explorations and realizations of  complex topological states of matter by engineering emergent symmetries in simple systems using judiciously chosen Floquet driving protocols.
 
We kindly acknowledge financial support by G. Anderheggen and the ETH Z\"{u}rich Foundation, and the productivity in research fellowship from CNPq. The authors would like to thank T. Pereg-Barnea, H.-Y. Kee, Y.-B. Kim, A. Schnyder, M. Rudner and T. Bzdu\vaccent{s}ek for fruitful discussions. 
 
\clearpage

\begin{widetext}

\section{Supplementary Material for: Generating quantum multi-criticality in topological insulators by periodic driving}
\beginsupplement

In the following we summarize some of the concepts and derivations used in the main text.

\section{Wannier state correlation function and critical exponents}
The critical exponents obtained from the CRG analysis can be related to those assigned to correlation length and susceptibility in the Landau theory of phase transitions.
This stems from that fact that the Fourier transform of the curvature function yields a \textit{Wannier state correlation function}, as we demonstrate below following the recipe in Ref.~[29] and [30] of the main text. For a 2D (stroboscopic) Hamiltonian described by (Floquet-)Bloch states $\left| u_{n \mathbf{k}} \right>$, such as the one considered in this work, the curvature function takes the form of a Berry curvature $F(\mathbf{k}, \mathbf{M}) = \sum_{n \in v} \nabla_{\mathbf{k}} \times \left< u_{n \mathbf{k}} \right| i \nabla_{\mathbf{k}} \left| u_{n \mathbf{k}} \right>$.
This expression can be equivalently expressed in terms of Wannier states $\left| \mathbf{R} n \right> = \frac{1}{N} \sum_{\mathbf{k}} e^{i \mathbf{k} \cdot \left( \hat{\mathbf{r}} - \mathbf{R} \right)} \left| u_{n \mathbf{k}} \right>$, where $N$ denotes the number of lattice sites, and $\hat{\mathbf{r}}$ is the position operator. 
We emphasize that in our analysis of the Floquet problem we work with the Floquet bands derived from the \textit{stroboscopic} effective Hamiltonian, and therefore the Wannier states calculated in this context are also to be interpreted as \textit{stroboscopic} states.
The wave function constructed from the Wannier states describe a real-space Wannier function $W_n (\mathbf{r} - \mathbf{R}) \equiv \left< \mathbf{r} \middle| \mathbf{R} n \right>$ at position $\mathbf{r}$ centered at the home cell $\mathbf{R}$.  
In this basis, the Berry curvature is written as~\cite{Wang:2006, Marzari:2012, Gradhand:2012} $F(\mathbf{k}, \mathbf{M}) = -i \sum_{n \in v} \sum_{\mathbf{R}} e^{-i \mathbf{k} \cdot \mathbf{R}} \left< \mathbf{R} n \right|  (\mathbf{R} \times \hat{\mathbf{r}})_z \left| \mathbf{0} n \right>$, where $\left| \mathbf{0} \right>$ denotes the Wannier function centered at the origin.
The Wannier representation allows us to draw a direct correspondence between the topological description of the system and the theory of orbital magnetization for 2D TRS-breaking systems~\cite{Thonhauser:2005,Xiao:2005,Ceresoli:2006,Shi:2007,Souza:2008}.
In this picture, the Fourier transform of the curvature function yields 
\begin{eqnarray}
&&\tilde{F}_{2D}(\mathbf{R}) = \frac{1}{(2\pi)^2} \iint \mathrm{d}^2 k \: e^{i \mathbf{k} \cdot \mathbf{R}}F(\mathbf{k},{\bf M}) = - i \sum_{n \in v} \left< \mathbf{R} n \right| \left(\mathbf{R} \times \mathbf{r} \right)_z \left| \mathbf{0} n \right> 
\nonumber \\
&&= -i \sum_{n \in v} \int \mathrm{d}^2 r W_n^*(\mathbf{r} - \mathbf{R}) \left( \mathbf{R} \times \hat{\mathbf{r}} \right)  W_n(\mathbf{r})
\propto e^{-R_x/\xi_{k_{x}}} e^{-R_y/\xi_{k_{y}}}\;,
\end{eqnarray}
which is a measure of the overlap of the Wannier function centered at $\mathbf{R}$ with that centered at the origin, sandwiched by the operator $(\mathbf{R} \times \mathbf{r})_z$. Note that the correlation function is a gauge-invariant observable because it is obtained upon integrating the gauge-invariant curvature function over a closed surface.

We can see that the correlation function $\tilde{F}_{2D}(\mathbf{R})$ decays exponentially with characteristic length scales $\xi_{i}$, indicating that $\xi_{i}$ acquire the meaning of correlation lengths of the topological phase transition with the associated critical exponents $\nu_i$. 
Furthermore, the integration of the correlation function over real space yields $\int \tilde{F}_{2D}(\mathbf{R}) \mathrm{d}^2 R=F({\mathbf k}_{0},{\bf M})$. 
The curvature function at the high symmetry point $F({\mathbf k}_{0},{\bf M})$ can be therefore interpreted as the analogue of the susceptibility in the Landau paradigm. For this reason, we assign to it the exponent $\gamma$ that characterizes its criticality. Consequently, different values of $\left\{\nu_{i},\gamma\right\}$ signify different universality classes. In Table I we summarize the critical exponents and the low energy dispersions of the two different universality classes obtained from the model in the main text. Details of fitting the critical exponents will be demonstrated in the following sections.
\begin{table}[h!]
	\centering
	\begin{tabular}{|c | c | c | c | c |}
	\hline
	topological phase transition& dispersion & $\gamma$ & $\nu_x$ or $\nu_{\tilde{x}}$ & $\nu_y$ \\
	\hline \hline
	$\tilde{J} = \frac{1}{3} \left(n - J \right)$ & $E \propto \sqrt{\tilde{\mathbf{k}}^2 + M^2}$ & $2$ & $1$ & $1$ \\
	$\tilde{J} = J$ & $E \propto \sqrt{(\tilde{k}_x^2 - \tilde{k}_y^2 )^2 + M^2}$ & $\frac{3}{2}$ & $\frac{1}{2}$ & $1$ \\	\hline 
	\end{tabular}
	\caption{Summary of the critical exponents extracted for the two different topological phase transitions existing in the Floquet-Chern insulator, where $\tilde{\mathbf{k}}$ refers to the $k$-coordinates around the HSPs where the corresponding gap closes.}
	\label{table_phases}
\end{table}

\section{Measures of topology for Floquet systems}

In this section we review how to construct topological invariants for 2D Floquet systems.
For a general time-periodic system with open boundary conditions described by the Hamiltonian $H(t)=H(t+T)$, the full dynamics of the topological edge states is governed by the time evolution operator, defined as $U(t) = \mathcal{T} \left\{ \exp \left[ -i \int_0^t \mathrm{d} t' \: H(t') \right] \right\}$, where $\mathcal{T}$ is the time-ordering operator and we have set the initial time $t_0=0$ and $\hbar=1$.
The operator $U(t)$ accounts for the full time dynamics, including the micromotion between periods.
When $t \to T$, the time evolution operator is typically called Floquet operator and,
because of the underlying time periodicity, it fulfills $U(T,0) = U(Tm,T(m-1))$ with $m \in \mathbb{N}$. 
This induces a discrete quantum map that describes stroboscopic dynamics~\cite{Haenggi}. 
We can then define an effective stroboscopic Floquet Hamiltonian via $U(T,0) \equiv e^{-i h_{\text{eff}} T}$ that contains the full information about the system at multiples of the driving period $T$.
Diagonalization of $h_{\text{eff}} T$ will then yield the stroboscopic quasienergy spectrum $\epsilon_{\alpha, k}$ of the Floquet-state solutions $\Psi_{\alpha}(k,t) = \exp(-i\epsilon_{\alpha,k} t) \Phi_{\alpha}(k,t)$, where $\Phi_{\alpha}(k,t) = \Phi_{\alpha}(k, t+ T)$~\citep{Haenggi}. 
Because of the $T$-periodicity of the Floquet modes $\Phi_{\alpha}(k,t)$, the quasienergies are defined modulo $\frac{2\pi}{T}=\omega$.
Therefore, we can restrict ourselves to consider a first ``{{Floquet-}}Brillouin zone'' of quasienergies $\epsilon_{\alpha} \in (-\omega/2, \omega/2)$. 
The number of stroboscopic edge modes can be determined as a function of the driving parameters from an analysis of the quasienergy spectrum, \textit{i.e.} from gap closures and localization of $0$ and $\pi$-quasienergy states.
Consequently, the topological phase diagram of the stroboscopic system can be ascertained.

In analogy with time-independent systems, the number of edge modes in an open geometry is related to the properties of the bulk time evolution operator via a bulk-edge correspondence~\cite{Rudner:2013}.
We could therefore determine the topological phase diagram by investigating the bulk time operator $
U(\mathbf{k},t) = \mathcal{T} \left\{ \exp \left[ -i \int_0^t \mathrm{d} t' \: H({\mathbf{k}},t') \right] \right\}$.
For a two-dimensional system like the one considered here, the number of edge modes at quasienergy $\epsilon$ can be calculated from a topological invariant defined as the winding number of an explicitly time-dependent map $S^1 \times S^1 \times S^1 \to U(N)$ constructed from the bulk-time operator~\cite{Bott:1978,Rudner:2013} $W[U_{\epsilon}] = \frac{1}{16 \pi^2} \int \mathrm{d}t \, \mathrm{d} k_x \, \mathrm{d} k_y  \Tr\left[ U_{\epsilon}^{-1} \partial_t U_{\epsilon} \left[ U_{\epsilon}^{-1} \partial_{k_x} U_{\epsilon}, U_{\epsilon}^{-1} \partial_{k_y} U_{\epsilon}\right]  \right]$.
Here, $U_{\epsilon}(\mathbf{k},t)$ is an operator derived from $U(\mathbf{k},t)$ preserving the number of edge modes at $\epsilon$ while smoothening the operator at the end of the cycle to the identity, \textit{i.e.} $U_{\epsilon}(\mathbf{k},T)=\mathds{1}$~\cite{Rudner:2013}. 
This transformation is necessary, because the winding number is equal to the number of edge modes at $\epsilon$ only if the spectrum of the bulk operator is gapped everywhere expect at $\epsilon$, which can be achieved only if $U_{\epsilon}(\mathbf{k},T)=\mathds{1}$. 

\section{Structure of the stroboscopic effective Hamiltonian}
In this section we elaborate on the structure of the low-energy theory of the topological phase transitions by explicitly looking at the form of the stroboscopic effective Hamiltonian and the curvature function given by Eq.~(2) and (3) in the main text. 
The stroboscopic quasienergy dispersion $\theta(\mathbf{k})$ corresponding to the effective Hamiltonian fulfills the eigenvalue equation~\cite{Molignini:2017}
\begin{equation}
h_{\text{eff}}(\mathbf{k}) \psi(\mathbf{k}) = \frac{\theta_{\mathbf{k}}}{T}  \psi(\mathbf{k})
\end{equation}
or, equivalently,
\begin{equation}
U_{\mathbf{k}}(T,0) \psi(\mathbf{k}) = e^{-i \theta_{\mathbf{k}}} \psi(\mathbf{k}).
\end{equation}
Hence, $\theta(\mathbf{k})$ can be derived, by calculating the eigenvalues of the Floquet operator and exploiting the identity $\arccos(z) = -i\log \left(z + \sqrt{z^2 - 1} \right)$, to be
\begin{equation}
\theta(\mathbf{k}) = -i \log \lambda_+ = \arccos \left[ \frac{\Tr U_{\mathbf{k}}(T,0)}{2} \right].
\end{equation}
The behavior of $\theta(\mathbf{k})$ at criticality can be used to shed light on the type of topological phase transition taking place there. 
Typically, the order of the gap closure is associated to the type of low energy theory.

\begin{figure}[h!]
\includegraphics[scale=1.0]{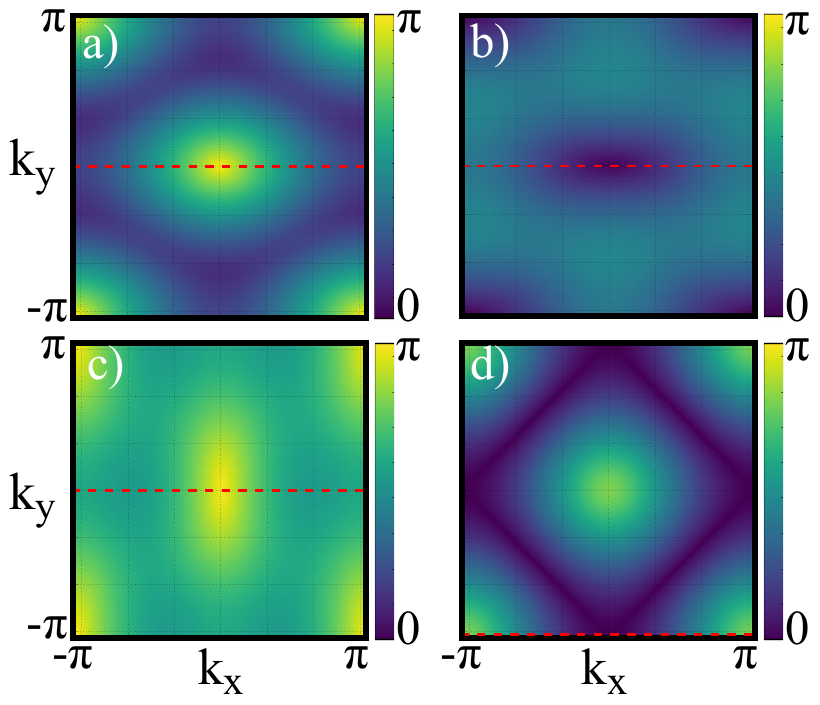}
\centering
\caption{Illustration of the gap closures of the quasienergy dispersion $\theta(\mathbf{k})$ at a) $J=0.1 [4\pi], \tilde{J}=0.3 [4\pi]$, b) $J=0.2 [4\pi], \tilde{J}=0.6 [4\pi]$, c) $J=0.3 [4\pi], \tilde{J}=0.9 [4\pi]$, d) $J=0.6 [4\pi], \tilde{J}=0.6 [4\pi]$. The dashed red lines indicate the cuts shown in Fig.~\ref{fig:gap-closures-cuts}}
\label{fig:gap-closures}
\end{figure}

For the four different transition lines in the phase diagram of the Floquet system analyzed in the main text, we show in Fig.~\ref{fig:gap-closures} contour plots of the quasienergy dispersion that illustrate the location of the gap closures.
Additionally, in Fig.~\ref{fig:gap-closures-cuts} we also present one-dimensional cuts that depict the form of the gap closures: we can appreciate that the topological phase transitions at $\tilde{J} = \frac{1}{3} (n  - J)$ are all characterized by linear gap closures, while the one at $\tilde{J} = J$ corresponds instead to a quadratic gap closure.
This visual finding was also verified through a numerical fit of the quasienergy dispersion.
\begin{figure}[h!]
\includegraphics[scale=1.0]{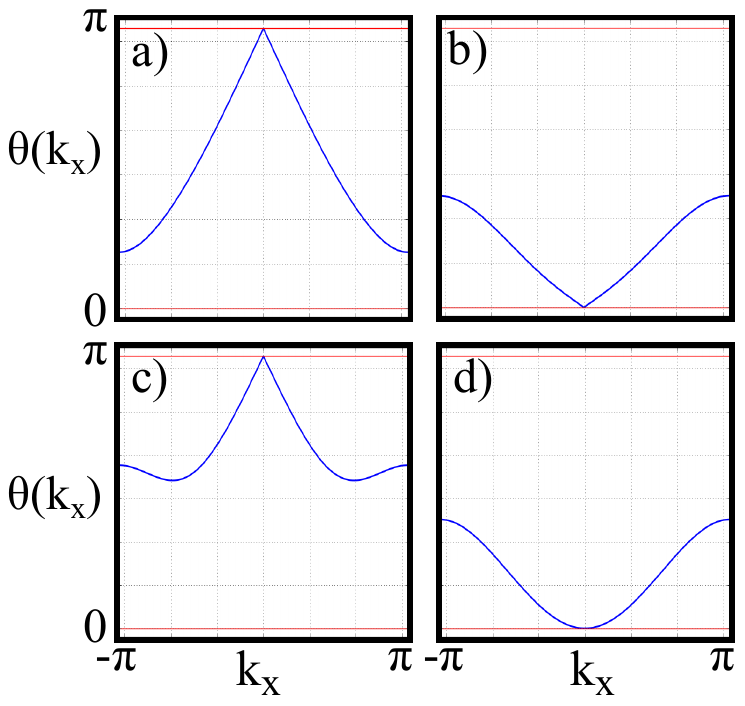}
\centering
\caption{Cuts of the quasienergy dispersion $\theta(k_x)$ along the $k_y$ value of the HSPs, illustrating the kind of gap closure for each topological phase transition. The values of the energy parameters are as in Fig.~\ref{fig:gap-closures}.}
\label{fig:gap-closures-cuts}
\end{figure}

\section{Comparison between Floquet-Chern insulator and nodal loop semimetal}
In this section we compare the behavior of the stroboscopic Berry curvature in Eq.~(3) of the main text for the Floquet-Chern insulator to the one of a static effective nodal loop semimetal and show that they exhibit the same behavior across the corresponding topological phase transitions.
For the Floquet-Chern insulator, we construct the Berry curvature from the effective stroboscopic Hamiltonian in Eq.~(2) of the main text.
To obtain the Berry curvature of the nodal loop semimetal, we use instead the static Hamiltonian
\begin{equation}
H_{NL} = \lambda_1 \sin k_x \sigma^x - \lambda_2 \sin k_y \sigma^y + (\mu - 2 (\cos k_x + \cos k_y)) \sigma^x,
\label{nodal-loop-Ham}
\end{equation}
which is a simplified version of the nodal-loop Hamiltonian presented in the main text. We have verified that both nodal-loop Hamiltonians reproduce the same behavior of the Berry curvature across the transition. The reason why we present the results from Hamiltonian \eqref{nodal-loop-Ham} is to draw a connection to Refs.~\cite{Qi:2006,Li:2016} in which it was already discussed.

A comparison of the two Berry curvatures across the topological phase transitions ($J=\tilde{J}=0.6$ for the Floquet-Chern insulator and correspondingly $\lambda_1 \simeq \lambda_2 = \mu$ for the nodal loop semimetal) is shown in Fig.~\ref{fig:nodal-loop} and reveals an excellent agreement between the two systems, indicating that the low-energy theory of the topological phase transition in the Floquet-Chern insulator is in fact a nodal loop semimetal.

As explained in the main text, the curved shape of the Berry curvature along the $k_x$-direction modifies the scaling law that captures the conservation of the Chern number. 
To recollect this scaling law, the correlation length must also be defined along the same curve, \textit{e.g.} as $\xi_{\mathbf{k}_0,\tilde{x}} \equiv \frac{2}{FWHM \left( \max_{k_y} \left\{ F(\mathbf{k}) \right\} \right)}$. 
Following this idea, we have additionally extracted the critical exponents from the Berry curvature by performing a numerical fit of the diverging quantities $F[\mathbf{k}_0]$, $\xi_{\tilde{x}}$, and $\xi_y$, shown in Fig.~\ref{fig:crit-exp}.
The exponents are $\gamma \approx 1.491$, $\nu_{\tilde{x}} \approx 0.499$, $\nu_y \approx 1.000$, which fulfill the (anisotropic) Dirac theory scaling law $\nu_{\tilde{x}} + \nu_y = \gamma$.

Finally, we note that in the Floquet-Chern insulator, all points along the transition line $\tilde{J}=J$, with the exception of the flat band points, are mapped to the same diagonal transition $\mu=\lambda_1=\lambda_2$ in the phase diagram of the static nodal loop semimetal.
We have confirmed this by performing similar fits at other values of $\tilde{J}=J$, and obtained comparable results as the one shown in Fig.~\ref{fig:crit-exp}.

\begin{figure}[t]
\centering
\includegraphics[scale=0.35]{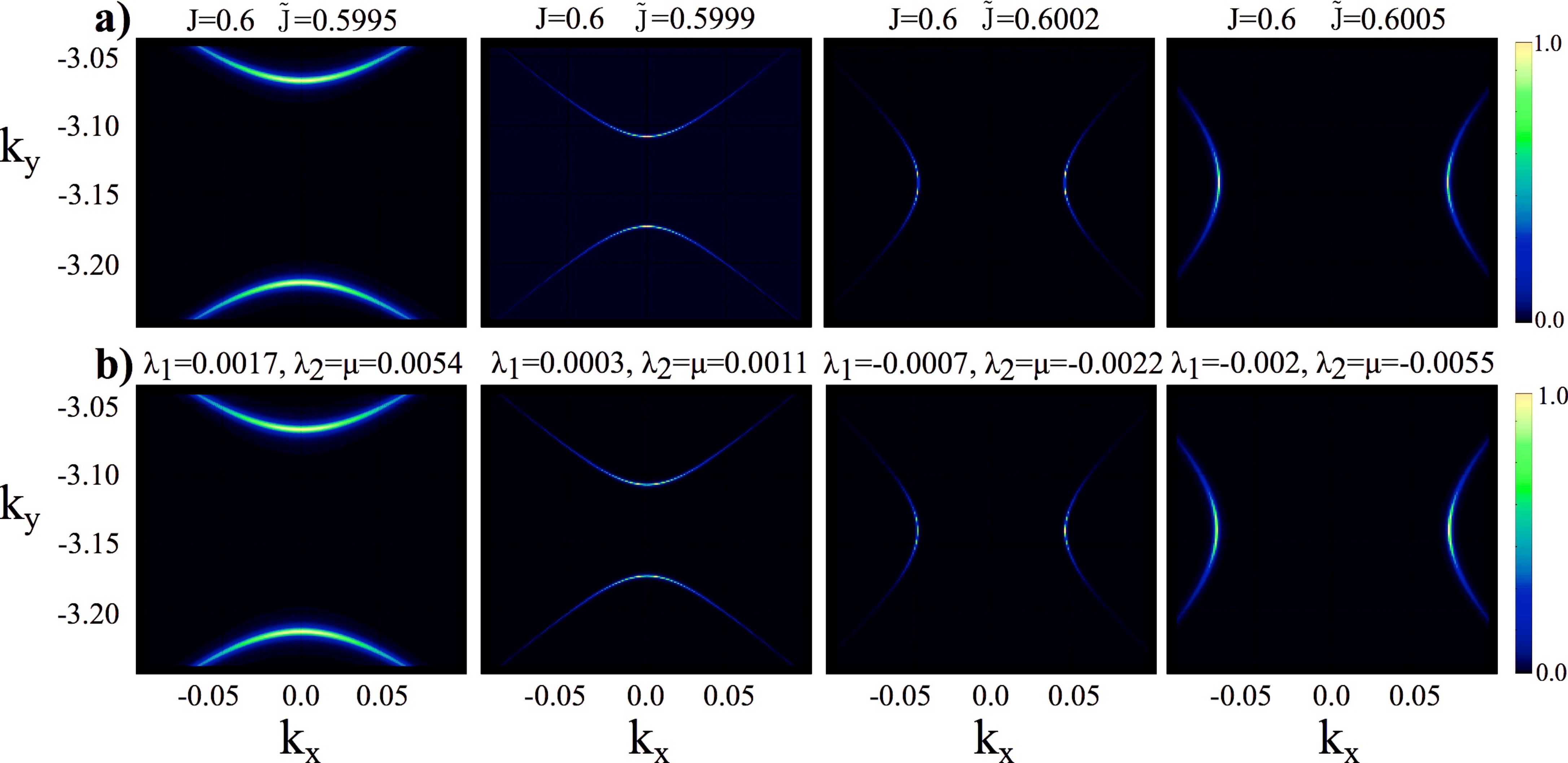}
\caption{
a) The Berry curvature across the topological phase transitionfor the Floquet-Chern insulator ($\tilde{J}$ in units of $4\pi$). 
b) The Berry curvature across the topological phase transitionfor the nodal loop semimetal.  
Note that for an easier comparison, $\lambda_1$ was chosen to be slightly different than $\lambda_2=\mu$.
Furthermore, the normalized absolute value of the Berry curvature was plotted.}
\label{fig:nodal-loop}
\end{figure}

\begin{figure}[t]
\centering
\includegraphics[scale=0.55]{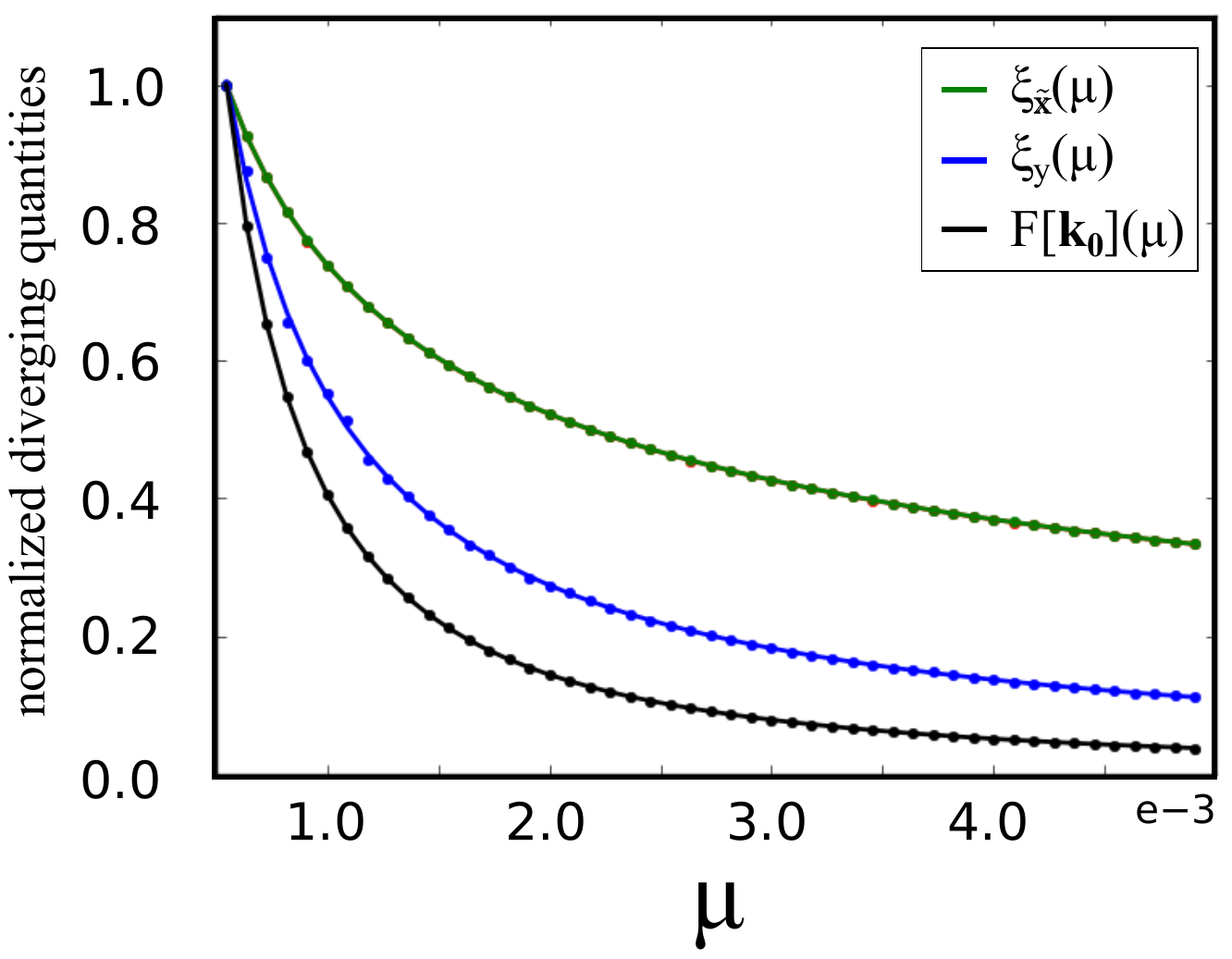}
\caption{Power law fit (solid lines) of the diverging quantities in the nodal loop semimetal extracted from the normalized Berry curvature. The fitted data is indicated by the dots.} 
\label{fig:crit-exp}
\end{figure}

\section{Emergent symmetry of the nodal loop semimetal}
We have numerically verified that the Floquet effective Hamiltonian fulfills the following symmetry operations:
\begin{align}
\mathcal{C} h_{\text{eff}}(k_x, k_y) \mathcal{C}^{-1} &= -h_{\text{eff}}(-k_x, -k_y)  &&\text{with} \quad \mathcal{C} = \sigma^z \circ \mathcal{K} \label{emp-symm-C} \\
\mathcal{I} h_{\text{eff}}(k_x, k_y) \mathcal{I}^{-1} &= h_{\text{eff}}(-k_x, -k_y)  &&\text{with} \quad \mathcal{I} = \sigma^x  \label{emp-symm-I} \\
\mathcal{D} h_{\text{eff}}(k_x, k_y) \mathcal{D}^{-1} &= h_{\text{eff}}(k_x \pm \pi, k_y \pm \pi)   &&\text{with} \quad \mathcal{D} = \sigma^z   \label{emp-symm-D}
\end{align}
These symmetries are charge conjugation, inversion, and displacement by $\pm \pi$ (reduction of the Brillouin zone to half).
Additionally, at $\tilde{J}=J$ corresponding to the nodal loop low-energy theory, a new symmetry emerges:
\begin{align}
\bar{\mathcal{M}} h_{\text{eff}}(k_x, k_y) \bar{\mathcal{M}}^{-1} &= - h_{\text{eff}}(k_y, k_x) &&\text{with} \quad \bar{\mathcal{M}} = \sigma^y. \label{chiral-mirror-symmetry}
\end{align}
We term this symmetry ``chiral mirror symmetry'', as it can be decomposed into a combination of a mirror symmetry $\mathcal{M}$ and chiral or sublattice symmetry $\mathcal{S}$. 
We note however that the two separate symmetry operations $\mathcal{M}$ and $\mathcal{S}$ need not be fulfilled when chiral mirror symmetry is present, much likely a chiral symmetry can exist on its own even when time-reversal and charge-conjugation symmetries are not separately fulfilled~\cite{ChiuReview:2016}.
In fact, we empirically verified that the effective Floquet Hamiltonian does not have separate mirror and chiral symmetries.

A possible nodal loop Hamiltonian that incorporates all the symmetries above is
\begin{align}
H_{NL} &= X(k_x, k_y) \sigma^x + Y(k_x, k_y) \sigma^y + Z(k_x, k_y) \sigma^z \\
&= (\mu - 2\eta(\cos k_x + \cos k_y))^2 \sigma^x + \alpha (\sin k_x - \sin k_y)( \cos k_x + \cos k_y) \sigma^y \nonumber \\
& \quad + \beta (\sin k_x +\sin k_y)( \cos k_x + \cos k_y) \sigma^z.
\label{nodal-loop-Ham-2}
\end{align}
The combination of charge conjugation and chiral mirror symmetry protects the nodal loop Hamiltonian \eqref{nodal-loop-Ham-2} from mass terms. 
Chiral mirror symmetry implies $Y(k_x, k_y) = - Y(k_y, k_x)$, which forbids mass terms in $\sigma^y$.
Charge conjugation instead implies $Z(k_x, k_y) = - Z(-k_x, -k_y)$, which forbids mass terms in $\sigma^z$.
The parameter $\mu$ controls the gap closure, such that the nodal loop sits at zero (quasi)energy when $\mu=0$ (and correspondingly $\tilde{J}=J$ for the Floquet-Chern insulator).

The chiral mirror symmetry is a direct consequence of the driving scheme applied to the Chern insulator and can be reformulated as a property of the Floquet operator $U(\mathbf{k},T)$:
\begin{align}
& \quad \qquad \bar{\mathcal{M}} h_{\text{eff}}(k_x, k_y) \bar{\mathcal{M}}^{-1} = - h_{\text{eff}}(k_y, k_x) \\
& \iff \qquad h^*_{\text{eff}}(k_x, k_y) = h_{\text{eff}}(k_y, k_x) \\
& \iff  \qquad \left( \log U(k_x, k_y; T) \right)^* = - \log U(k_y, k_x; T) \\
& \iff \qquad  \log U^*(k_x, k_y; T) = \log U^{-1}(k_y, k_x; T) \\
& \iff \qquad U(k_x, k_y; T) =  U^{\mathbb{T}}(k_y, k_x; T) 
\end{align}
This property can be generalized to other forms of mirror symmetries, such as reflections with respect to a different axis ($k_x \leftrightarrow k_y$ is a reflection with respect to $\hat{a} = \frac{1}{\sqrt{2}}(1,1)$ in $\mathbf{k}$-space). The general condition to have chiral mirror symmetry in the effective theory of the Floquet Hamiltonian is then 
\begin{equation}
U(\mathbf{k}, T) = U^{\mathbb{T}}(\mathbf{k}^{\mathcal{M}}, T)
\end{equation}
where $\mathbf{k}^{\mathcal{M}}$ are the $k$-coordinates transformed under the chiral mirror symmetry and $\mathbb{T}$ is matrix transposition.

\end{widetext}

\bibliography{many-body-biblio}

\begin{thebibliography}{70}
\expandafter\ifx\csname natexlab\endcsname\relax\def\natexlab#1{#1}\fi
\expandafter\ifx\csname bibnamefont\endcsname\relax
  \def\bibnamefont#1{#1}\fi
\expandafter\ifx\csname bibfnamefont\endcsname\relax
  \def\bibfnamefont#1{#1}\fi
\expandafter\ifx\csname citenamefont\endcsname\relax
  \def\citenamefont#1{#1}\fi
\expandafter\ifx\csname url\endcsname\relax
  \def\url#1{\texttt{#1}}\fi
\expandafter\ifx\csname urlprefix\endcsname\relax\def\urlprefix{URL }\fi
\providecommand{\bibinfo}[2]{#2}
\providecommand{\eprint}[2][]{\url{#2}}

\bibitem[{\citenamefont{Kitagawa et~al.}(2010)\citenamefont{Kitagawa, Berg,
  Rudner, and Demler}}]{Kitagawa}
\bibinfo{author}{\bibfnamefont{T.}~\bibnamefont{Kitagawa}},
  \bibinfo{author}{\bibfnamefont{E.}~\bibnamefont{Berg}},
  \bibinfo{author}{\bibfnamefont{M.}~\bibnamefont{Rudner}}, \bibnamefont{and}
  \bibinfo{author}{\bibfnamefont{E.}~\bibnamefont{Demler}},
  \bibinfo{journal}{Phys. Rev. B} \textbf{\bibinfo{volume}{82}},
  \bibinfo{pages}{235114} (\bibinfo{year}{2010}).

\bibitem[{\citenamefont{Lindner et~al.}(2011)\citenamefont{Lindner, Refael, and
  Galitski}}]{Lindner}
\bibinfo{author}{\bibfnamefont{N.~H.} \bibnamefont{Lindner}},
  \bibinfo{author}{\bibfnamefont{G.}~\bibnamefont{Refael}}, \bibnamefont{and}
  \bibinfo{author}{\bibfnamefont{V.}~\bibnamefont{Galitski}},
  \bibinfo{journal}{Nat. Phys.} \textbf{\bibinfo{volume}{7}},
  \bibinfo{pages}{490} (\bibinfo{year}{2011}).

\bibitem[{\citenamefont{Cayssol et~al.}(2013)\citenamefont{Cayssol, D{\'o}ra,
  Simon, and Moessner}}]{Cayssol:2013}
\bibinfo{author}{\bibfnamefont{J.}~\bibnamefont{Cayssol}},
  \bibinfo{author}{\bibfnamefont{B.}~\bibnamefont{D{\'o}ra}},
  \bibinfo{author}{\bibfnamefont{F.}~\bibnamefont{Simon}}, \bibnamefont{and}
  \bibinfo{author}{\bibfnamefont{R.}~\bibnamefont{Moessner}},
  \bibinfo{journal}{Phys. Status Solidi RRL} \textbf{\bibinfo{volume}{7}},
  \bibinfo{pages}{101} (\bibinfo{year}{2013}).

\bibitem[{\citenamefont{Harper and Roy}(2017)}]{Harper:2017}
\bibinfo{author}{\bibfnamefont{F.}~\bibnamefont{Harper}} \bibnamefont{and}
  \bibinfo{author}{\bibfnamefont{R.}~\bibnamefont{Roy}},
  \bibinfo{journal}{Phys. Rev. Lett.} \textbf{\bibinfo{volume}{118}},
  \bibinfo{pages}{115301} (\bibinfo{year}{2017}).

\bibitem[{\citenamefont{Roy and Harper}(2017)}]{Roy:2017}
\bibinfo{author}{\bibfnamefont{R.}~\bibnamefont{Roy}} \bibnamefont{and}
  \bibinfo{author}{\bibfnamefont{F.}~\bibnamefont{Harper}},
  \bibinfo{journal}{Phys. Rev. B} \textbf{\bibinfo{volume}{96}},
  \bibinfo{pages}{155118} (\bibinfo{year}{2017}).

\bibitem[{\citenamefont{Esin et~al.}(2018)\citenamefont{Esin, Rudner, Refael,
  and Lindner}}]{Esin:2018}
\bibinfo{author}{\bibfnamefont{I.}~\bibnamefont{Esin}},
  \bibinfo{author}{\bibfnamefont{M.~S.} \bibnamefont{Rudner}},
  \bibinfo{author}{\bibfnamefont{G.}~\bibnamefont{Refael}}, \bibnamefont{and}
  \bibinfo{author}{\bibfnamefont{N.~H.} \bibnamefont{Lindner}},
  \bibinfo{journal}{Phys. Rev. B} p. \bibinfo{pages}{245401}
  (\bibinfo{year}{2018}).

\bibitem[{\citenamefont{Liu et~al.}(2013)\citenamefont{Liu, Levchenko, and
  Baranger}}]{Liu}
\bibinfo{author}{\bibfnamefont{D.~E.} \bibnamefont{Liu}},
  \bibinfo{author}{\bibfnamefont{A.}~\bibnamefont{Levchenko}},
  \bibnamefont{and} \bibinfo{author}{\bibfnamefont{H.~U.}
  \bibnamefont{Baranger}}, \bibinfo{journal}{Phys. Rev. Lett.}
  \textbf{\bibinfo{volume}{111}}, \bibinfo{pages}{047002}
  (\bibinfo{year}{2013}).

\bibitem[{\citenamefont{Thakurathi et~al.}(2013)\citenamefont{Thakurathi,
  Patel, Sen, and Dutta}}]{Thakurathi}
\bibinfo{author}{\bibfnamefont{M.}~\bibnamefont{Thakurathi}},
  \bibinfo{author}{\bibfnamefont{A.~A.} \bibnamefont{Patel}},
  \bibinfo{author}{\bibfnamefont{D.}~\bibnamefont{Sen}}, \bibnamefont{and}
  \bibinfo{author}{\bibfnamefont{A.}~\bibnamefont{Dutta}},
  \bibinfo{journal}{Phys. Rev. B} \textbf{\bibinfo{volume}{88}},
  \bibinfo{pages}{155133} (\bibinfo{year}{2013}).

\bibitem[{\citenamefont{Thakurathi et~al.}(2014)\citenamefont{Thakurathi,
  Sengupta, and Sen}}]{Thakurathi:2014}
\bibinfo{author}{\bibfnamefont{M.}~\bibnamefont{Thakurathi}},
  \bibinfo{author}{\bibfnamefont{K.}~\bibnamefont{Sengupta}}, \bibnamefont{and}
  \bibinfo{author}{\bibfnamefont{D.}~\bibnamefont{Sen}},
  \bibinfo{journal}{Phys. Rev. B} p. \bibinfo{pages}{235434}
  (\bibinfo{year}{2014}).

\bibitem[{\citenamefont{Wang et~al.}(2014)\citenamefont{Wang, Q.-F.-Sun, and
  Xie}}]{Wang:2014}
\bibinfo{author}{\bibfnamefont{P.}~\bibnamefont{Wang}},
  \bibinfo{author}{\bibnamefont{Q.-F.-Sun}}, \bibnamefont{and}
  \bibinfo{author}{\bibfnamefont{X.~C.} \bibnamefont{Xie}},
  \bibinfo{journal}{Phys. Rev. B} \textbf{\bibinfo{volume}{90}},
  \bibinfo{pages}{155407} (\bibinfo{year}{2014}).

\bibitem[{\citenamefont{Sacramento}(2015)}]{Sacramento:2015}
\bibinfo{author}{\bibfnamefont{P.~D.} \bibnamefont{Sacramento}},
  \bibinfo{journal}{Phys. Rev. B} \textbf{\bibinfo{volume}{91}},
  \bibinfo{pages}{214518} (\bibinfo{year}{2015}).

\bibitem[{\citenamefont{Thakurathi et~al.}(2017)\citenamefont{Thakurathi, Loss,
  and Klinovaja}}]{Thakurathi:2017}
\bibinfo{author}{\bibfnamefont{M.}~\bibnamefont{Thakurathi}},
  \bibinfo{author}{\bibfnamefont{D.}~\bibnamefont{Loss}}, \bibnamefont{and}
  \bibinfo{author}{\bibfnamefont{J.}~\bibnamefont{Klinovaja}},
  \bibinfo{journal}{Phys. Rev. B} \textbf{\bibinfo{volume}{95}},
  \bibinfo{pages}{155407} (\bibinfo{year}{2017}).

\bibitem[{\citenamefont{Molignini et~al.}(2017)\citenamefont{Molignini, van
  Nieuwenburg, and Chitra}}]{Molignini:2017}
\bibinfo{author}{\bibfnamefont{P.}~\bibnamefont{Molignini}},
  \bibinfo{author}{\bibfnamefont{E.}~\bibnamefont{van Nieuwenburg}},
  \bibnamefont{and} \bibinfo{author}{\bibfnamefont{R.}~\bibnamefont{Chitra}},
  \bibinfo{journal}{Phys. Rev. B} \textbf{\bibinfo{volume}{96}},
  \bibinfo{pages}{125144} (\bibinfo{year}{2017}).

\bibitem[{\citenamefont{Molignini et~al.}(2018)\citenamefont{Molignini, Chen,
  and Chitra}}]{Molignini:2018}
\bibinfo{author}{\bibfnamefont{P.}~\bibnamefont{Molignini}},
  \bibinfo{author}{\bibfnamefont{W.}~\bibnamefont{Chen}}, \bibnamefont{and}
  \bibinfo{author}{\bibfnamefont{R.}~\bibnamefont{Chitra}},
  \bibinfo{journal}{Phys. Rev. B} \textbf{\bibinfo{volume}{98}},
  \bibinfo{pages}{125129} (\bibinfo{year}{2018}).

\bibitem[{\citenamefont{{\v C}ade{\v z} et~al.}(2019)\citenamefont{{\v C}ade{\v
  z}, Mondaini, and Sacramento}}]{Cadez:2019}
\bibinfo{author}{\bibfnamefont{T.}~\bibnamefont{{\v C}ade{\v z}}},
  \bibinfo{author}{\bibfnamefont{R.}~\bibnamefont{Mondaini}}, \bibnamefont{and}
  \bibinfo{author}{\bibfnamefont{P.~D.} \bibnamefont{Sacramento}},
  \bibinfo{journal}{Phys. Rev. B} \textbf{\bibinfo{volume}{99}},
  \bibinfo{pages}{014301} (\bibinfo{year}{2019}).

\bibitem[{\citenamefont{Bomantara et~al.}(2016)\citenamefont{Bomantara,
  Raghava, Zhou, and Gong}}]{Gong:2016}
\bibinfo{author}{\bibfnamefont{R.~W.} \bibnamefont{Bomantara}},
  \bibinfo{author}{\bibfnamefont{G.~N.} \bibnamefont{Raghava}},
  \bibinfo{author}{\bibfnamefont{L.}~\bibnamefont{Zhou}}, \bibnamefont{and}
  \bibinfo{author}{\bibfnamefont{J.}~\bibnamefont{Gong}},
  \bibinfo{journal}{Phys. Rev. E 93} \textbf{\bibinfo{volume}{93}},
  \bibinfo{pages}{022209} (\bibinfo{year}{2016}).

\bibitem[{\citenamefont{Zhou et~al.}(2016)\citenamefont{Zhou, Chen, and
  Gong}}]{Gong:2016-2}
\bibinfo{author}{\bibfnamefont{L.}~\bibnamefont{Zhou}},
  \bibinfo{author}{\bibfnamefont{C.}~\bibnamefont{Chen}}, \bibnamefont{and}
  \bibinfo{author}{\bibfnamefont{J.}~\bibnamefont{Gong}},
  \bibinfo{journal}{Phys. Rev. B} \textbf{\bibinfo{volume}{94}},
  \bibinfo{pages}{075443} (\bibinfo{year}{2016}).

\bibitem[{\citenamefont{Bomantara and Gong}(2016)}]{Gong:2016-3}
\bibinfo{author}{\bibfnamefont{R.~W.} \bibnamefont{Bomantara}}
  \bibnamefont{and} \bibinfo{author}{\bibfnamefont{J.}~\bibnamefont{Gong}},
  \bibinfo{journal}{Phys. Rev. B} \textbf{\bibinfo{volume}{94}},
  \bibinfo{pages}{235447} (\bibinfo{year}{2016}).

\bibitem[{\citenamefont{Bucciantini et~al.}(2017)\citenamefont{Bucciantini,
  Roy, Kitamura, and Oka}}]{Bucciantini:2017}
\bibinfo{author}{\bibfnamefont{L.}~\bibnamefont{Bucciantini}},
  \bibinfo{author}{\bibfnamefont{S.}~\bibnamefont{Roy}},
  \bibinfo{author}{\bibfnamefont{S.}~\bibnamefont{Kitamura}}, \bibnamefont{and}
  \bibinfo{author}{\bibfnamefont{T.}~\bibnamefont{Oka}},
  \bibinfo{journal}{Phys. Rev. B} \textbf{\bibinfo{volume}{96}},
  \bibinfo{pages}{041126(R)} (\bibinfo{year}{2017}).

\bibitem[{\citenamefont{H{\"u}bener et~al.}(2017)\citenamefont{H{\"u}bener,
  Sentef, Giovannini, Kemper, and Rubio}}]{Huebener:2017}
\bibinfo{author}{\bibfnamefont{H.}~\bibnamefont{H{\"u}bener}},
  \bibinfo{author}{\bibfnamefont{M.~A.} \bibnamefont{Sentef}},
  \bibinfo{author}{\bibfnamefont{U.~D.} \bibnamefont{Giovannini}},
  \bibinfo{author}{\bibfnamefont{A.~F.} \bibnamefont{Kemper}},
  \bibnamefont{and} \bibinfo{author}{\bibfnamefont{A.}~\bibnamefont{Rubio}},
  \bibinfo{journal}{Nat. Comm.} \textbf{\bibinfo{volume}{8}},
  \bibinfo{pages}{13940} (\bibinfo{year}{2017}).

\bibitem[{\citenamefont{Cao et~al.}(2017)\citenamefont{Cao, Qi, and
  Xiang}}]{Cao:2017}
\bibinfo{author}{\bibfnamefont{J.}~\bibnamefont{Cao}},
  \bibinfo{author}{\bibfnamefont{F.}~\bibnamefont{Qi}}, \bibnamefont{and}
  \bibinfo{author}{\bibfnamefont{Y.}~\bibnamefont{Xiang}},
  \bibinfo{journal}{Europhys. Lett.} \textbf{\bibinfo{volume}{119}},
  \bibinfo{pages}{57008} (\bibinfo{year}{2017}).

\bibitem[{\citenamefont{Li et~al.}(2017)\citenamefont{Li, Chesi, Yin, and
  Chen}}]{ShuChen:2017}
\bibinfo{author}{\bibfnamefont{L.}~\bibnamefont{Li}},
  \bibinfo{author}{\bibfnamefont{S.}~\bibnamefont{Chesi}},
  \bibinfo{author}{\bibfnamefont{C.}~\bibnamefont{Yin}}, \bibnamefont{and}
  \bibinfo{author}{\bibfnamefont{S.}~\bibnamefont{Chen}},
  \bibinfo{journal}{Phys. Rev. B} \textbf{\bibinfo{volume}{96}},
  \bibinfo{pages}{081116(R)} (\bibinfo{year}{2017}).

\bibitem[{\citenamefont{Chen et~al.}(2018)\citenamefont{Chen, Zhou, and
  Xu}}]{ChenZhou:2018}
\bibinfo{author}{\bibfnamefont{R.}~\bibnamefont{Chen}},
  \bibinfo{author}{\bibfnamefont{B.}~\bibnamefont{Zhou}}, \bibnamefont{and}
  \bibinfo{author}{\bibfnamefont{D.-H.} \bibnamefont{Xu}},
  \bibinfo{journal}{Phys. Rev. B} \textbf{\bibinfo{volume}{97}},
  \bibinfo{pages}{155152} (\bibinfo{year}{2018}).

\bibitem[{\citenamefont{Li et~al.}(2018)\citenamefont{Li, Lee, and
  Gong}}]{Li:2018}
\bibinfo{author}{\bibfnamefont{L.}~\bibnamefont{Li}},
  \bibinfo{author}{\bibfnamefont{C.~H.} \bibnamefont{Lee}}, \bibnamefont{and}
  \bibinfo{author}{\bibfnamefont{J.}~\bibnamefont{Gong}},
  \bibinfo{journal}{Phys. Rev. Lett.} \textbf{\bibinfo{volume}{121}},
  \bibinfo{pages}{036401} (\bibinfo{year}{2018}).

\bibitem[{\citenamefont{Fisher}(1984)}]{multicriticality-book}
\bibinfo{author}{\bibfnamefont{M.}~\bibnamefont{Fisher}},
  \emph{\bibinfo{title}{Multicriticality: A Theoretical Introduction. In: Pynn
  R., Skjeltorp A. (eds) Multicritical Phenomena}}
  (\bibinfo{publisher}{Springer}, \bibinfo{address}{Boston, MA},
  \bibinfo{year}{1984}).

\bibitem[{\citenamefont{Zacharias et~al.}(2009)\citenamefont{Zacharias,
  W\"{o}lfle, and Garst}}]{Zacharias:2009}
\bibinfo{author}{\bibfnamefont{M.}~\bibnamefont{Zacharias}},
  \bibinfo{author}{\bibfnamefont{P.}~\bibnamefont{W\"{o}lfle}},
  \bibnamefont{and} \bibinfo{author}{\bibfnamefont{M.}~\bibnamefont{Garst}},
  \bibinfo{journal}{Phys. Rev. B} \textbf{\bibinfo{volume}{80}},
  \bibinfo{pages}{165116} (\bibinfo{year}{2009}).

\bibitem[{\citenamefont{Carr}(2010)}]{Carr-book}
\bibinfo{author}{\bibfnamefont{L.}~\bibnamefont{Carr}},
  \emph{\bibinfo{title}{Understanding Quantum Phase Transitions}}, Condensed
  Matter Physics (\bibinfo{publisher}{CRC Press}, \bibinfo{year}{2010}).

\bibitem[{\citenamefont{Pixley et~al.}(2014)\citenamefont{Pixley, Shashi, and
  Nevidomsky}}]{Pixley:2014}
\bibinfo{author}{\bibfnamefont{J.~H.} \bibnamefont{Pixley}},
  \bibinfo{author}{\bibfnamefont{A.}~\bibnamefont{Shashi}}, \bibnamefont{and}
  \bibinfo{author}{\bibfnamefont{A.~H.} \bibnamefont{Nevidomsky}},
  \bibinfo{journal}{Phys. Rev. B} \textbf{\bibinfo{volume}{90}},
  \bibinfo{pages}{214426} (\bibinfo{year}{2014}).

\bibitem[{\citenamefont{Brando et~al.}(2016)\citenamefont{Brando, Kerkau,
  Todorova, Yamada, Khuntia, F\"{o}rster, Burkhard, Baenitz, and
  Kreiner}}]{Brando:2016}
\bibinfo{author}{\bibfnamefont{M.}~\bibnamefont{Brando}},
  \bibinfo{author}{\bibfnamefont{A.}~\bibnamefont{Kerkau}},
  \bibinfo{author}{\bibfnamefont{A.}~\bibnamefont{Todorova}},
  \bibinfo{author}{\bibfnamefont{Y.}~\bibnamefont{Yamada}},
  \bibinfo{author}{\bibfnamefont{P.}~\bibnamefont{Khuntia}},
  \bibinfo{author}{\bibfnamefont{T.}~\bibnamefont{F\"{o}rster}},
  \bibinfo{author}{\bibfnamefont{U.}~\bibnamefont{Burkhard}},
  \bibinfo{author}{\bibfnamefont{M.}~\bibnamefont{Baenitz}}, \bibnamefont{and}
  \bibinfo{author}{\bibfnamefont{G.}~\bibnamefont{Kreiner}},
  \bibinfo{journal}{J. Phys. Soc. Jpn.} \textbf{\bibinfo{volume}{85}},
  \bibinfo{pages}{084707} (\bibinfo{year}{2016}).

\bibitem[{\citenamefont{Chen}(2016)}]{Chen:2016}
\bibinfo{author}{\bibfnamefont{W.}~\bibnamefont{Chen}}, \bibinfo{journal}{J.
  Phys.: Condens. Matter} \textbf{\bibinfo{volume}{28}},
  \bibinfo{pages}{055601} (\bibinfo{year}{2016}).

\bibitem[{\citenamefont{Chen et~al.}(2016)\citenamefont{Chen, Sigrist, and
  Schnyder}}]{Chen-Sigrist:2016}
\bibinfo{author}{\bibfnamefont{W.}~\bibnamefont{Chen}},
  \bibinfo{author}{\bibfnamefont{M.}~\bibnamefont{Sigrist}}, \bibnamefont{and}
  \bibinfo{author}{\bibfnamefont{A.~P.} \bibnamefont{Schnyder}},
  \bibinfo{journal}{J. Phys.: Condens. Matter} \textbf{\bibinfo{volume}{28}},
  \bibinfo{pages}{365501} (\bibinfo{year}{2016}).

\bibitem[{\citenamefont{Kourtis et~al.}(2017)\citenamefont{Kourtis, Neupert,
  Mudry, Sigrist, and Chen}}]{Kourtis:2017}
\bibinfo{author}{\bibfnamefont{S.}~\bibnamefont{Kourtis}},
  \bibinfo{author}{\bibfnamefont{T.}~\bibnamefont{Neupert}},
  \bibinfo{author}{\bibfnamefont{C.}~\bibnamefont{Mudry}},
  \bibinfo{author}{\bibfnamefont{M.}~\bibnamefont{Sigrist}}, \bibnamefont{and}
  \bibinfo{author}{\bibfnamefont{W.}~\bibnamefont{Chen}},
  \bibinfo{journal}{Phys. Rev. B} \textbf{\bibinfo{volume}{96}},
  \bibinfo{pages}{205117} (\bibinfo{year}{2017}).

\bibitem[{\citenamefont{Chen}(2018)}]{Chen:2018}
\bibinfo{author}{\bibfnamefont{W.}~\bibnamefont{Chen}}, \bibinfo{journal}{Phys.
  Rev. B} \textbf{\bibinfo{volume}{97}}, \bibinfo{pages}{115130}
  (\bibinfo{year}{2018}).

\bibitem[{\citenamefont{Rufo et~al.}(2019)\citenamefont{Rufo, Lopes,
  Continentino, and Griffith}}]{Rufo:2019}
\bibinfo{author}{\bibfnamefont{S.}~\bibnamefont{Rufo}},
  \bibinfo{author}{\bibfnamefont{N.}~\bibnamefont{Lopes}},
  \bibinfo{author}{\bibfnamefont{M.~A.} \bibnamefont{Continentino}},
  \bibnamefont{and} \bibinfo{author}{\bibfnamefont{M.~A.~R.}
  \bibnamefont{Griffith}}, \bibinfo{journal}{arXiv:1907.12701}
  (\bibinfo{year}{2019}).

\bibitem[{\citenamefont{Rudner et~al.}(2013)\citenamefont{Rudner, Lindner,
  Berg, and Levin}}]{Rudner:2013}
\bibinfo{author}{\bibfnamefont{M.~S.} \bibnamefont{Rudner}},
  \bibinfo{author}{\bibfnamefont{N.~H.} \bibnamefont{Lindner}},
  \bibinfo{author}{\bibfnamefont{E.}~\bibnamefont{Berg}}, \bibnamefont{and}
  \bibinfo{author}{\bibfnamefont{M.}~\bibnamefont{Levin}},
  \bibinfo{journal}{Phys. Rev. X} \textbf{\bibinfo{volume}{3}},
  \bibinfo{pages}{031005} (\bibinfo{year}{2013}).

\bibitem[{\citenamefont{Mukherjee et~al.}(2017)\citenamefont{Mukherjee,
  Spracklen, Valiente, Andersson, \"{O}hberg, Goldman, and
  Thomson}}]{Mukherjee:2017}
\bibinfo{author}{\bibfnamefont{S.}~\bibnamefont{Mukherjee}},
  \bibinfo{author}{\bibfnamefont{A.}~\bibnamefont{Spracklen}},
  \bibinfo{author}{\bibfnamefont{M.}~\bibnamefont{Valiente}},
  \bibinfo{author}{\bibfnamefont{E.}~\bibnamefont{Andersson}},
  \bibinfo{author}{\bibfnamefont{P.}~\bibnamefont{\"{O}hberg}},
  \bibinfo{author}{\bibfnamefont{N.}~\bibnamefont{Goldman}}, \bibnamefont{and}
  \bibinfo{author}{\bibfnamefont{R.~R.} \bibnamefont{Thomson}},
  \bibinfo{journal}{Nature Communications} \textbf{\bibinfo{volume}{8}}
  (\bibinfo{year}{2017}).

\bibitem[{\citenamefont{Yao et~al.}(2017)\citenamefont{Yao, Yan, and
  Wang}}]{Yao:2017}
\bibinfo{author}{\bibfnamefont{S.}~\bibnamefont{Yao}},
  \bibinfo{author}{\bibfnamefont{Z.}~\bibnamefont{Yan}}, \bibnamefont{and}
  \bibinfo{author}{\bibfnamefont{Z.}~\bibnamefont{Wang}},
  \bibinfo{journal}{Phys. Rev. B} \textbf{\bibinfo{volume}{96}},
  \bibinfo{pages}{195303} (\bibinfo{year}{2017}).

\bibitem[{\citenamefont{Qi et~al.}(2006)\citenamefont{Qi, Wu, and
  Zhang}}]{Qi:2006}
\bibinfo{author}{\bibfnamefont{X.-L.} \bibnamefont{Qi}},
  \bibinfo{author}{\bibfnamefont{Y.-S.} \bibnamefont{Wu}}, \bibnamefont{and}
  \bibinfo{author}{\bibfnamefont{S.-C.} \bibnamefont{Zhang}},
  \bibinfo{journal}{Phys. Rev. B} \textbf{\bibinfo{volume}{74}},
  \bibinfo{pages}{085308} (\bibinfo{year}{2006}).

\bibitem[{\citenamefont{Li and Ara{\'u}jo}(2017)}]{Li:2016}
\bibinfo{author}{\bibfnamefont{L.}~\bibnamefont{Li}} \bibnamefont{and}
  \bibinfo{author}{\bibfnamefont{M.~A.~N.} \bibnamefont{Ara{\'u}jo}},
  \bibinfo{journal}{Phys. Rev. B} \textbf{\bibinfo{volume}{94}},
  \bibinfo{pages}{165117} (\bibinfo{year}{2017}).

\bibitem[{\citenamefont{\"{U}nal
  et~al.}(2019{\natexlab{a}})\citenamefont{\"{U}nal, Seradjeh, and
  Eckardt}}]{Unal:2019}
\bibinfo{author}{\bibfnamefont{F.~N.} \bibnamefont{\"{U}nal}},
  \bibinfo{author}{\bibfnamefont{B.}~\bibnamefont{Seradjeh}}, \bibnamefont{and}
  \bibinfo{author}{\bibfnamefont{A.}~\bibnamefont{Eckardt}},
  \bibinfo{journal}{Phys. Rev. Lett.} \textbf{\bibinfo{volume}{122}},
  \bibinfo{pages}{253601} (\bibinfo{year}{2019}{\natexlab{a}}).

\bibitem[{\citenamefont{\"{U}nal
  et~al.}(2019{\natexlab{b}})\citenamefont{\"{U}nal, Eckardt, and
  Slager}}]{Unal:2019-2}
\bibinfo{author}{\bibfnamefont{F.~N.} \bibnamefont{\"{U}nal}},
  \bibinfo{author}{\bibfnamefont{A.}~\bibnamefont{Eckardt}}, \bibnamefont{and}
  \bibinfo{author}{\bibfnamefont{R.-J.} \bibnamefont{Slager}},
  \bibinfo{journal}{arXiv:1904.03202}  (\bibinfo{year}{2019}{\natexlab{b}}).

\bibitem[{sup()}]{supmat}
\emph{\bibinfo{title}{Supplementary material}}.

\bibitem[{\citenamefont{Chiu et~al.}(2016)\citenamefont{Chiu, Teo, Schnyder,
  and Ryu}}]{ChiuReview:2016}
\bibinfo{author}{\bibfnamefont{C.-K.} \bibnamefont{Chiu}},
  \bibinfo{author}{\bibfnamefont{J.~C.~Y.} \bibnamefont{Teo}},
  \bibinfo{author}{\bibfnamefont{A.~P.} \bibnamefont{Schnyder}},
  \bibnamefont{and} \bibinfo{author}{\bibfnamefont{S.}~\bibnamefont{Ryu}},
  \bibinfo{journal}{Rev. Mod. Phys.} \textbf{\bibinfo{volume}{88}},
  \bibinfo{pages}{035055} (\bibinfo{year}{2016}).

\bibitem[{\citenamefont{Chen et~al.}(2017)\citenamefont{Chen, Legner,
  R\"{u}egg, and Sigrist}}]{Chen:2017}
\bibinfo{author}{\bibfnamefont{W.}~\bibnamefont{Chen}},
  \bibinfo{author}{\bibfnamefont{M.}~\bibnamefont{Legner}},
  \bibinfo{author}{\bibfnamefont{A.}~\bibnamefont{R\"{u}egg}},
  \bibnamefont{and} \bibinfo{author}{\bibfnamefont{M.}~\bibnamefont{Sigrist}},
  \bibinfo{journal}{Phys. Rev. B} \textbf{\bibinfo{volume}{95}},
  \bibinfo{pages}{075116} (\bibinfo{year}{2017}).

\bibitem[{\citenamefont{Chen and Schnyder}(2019)}]{Chen-Schnyder:2019}
\bibinfo{author}{\bibfnamefont{W.}~\bibnamefont{Chen}} \bibnamefont{and}
  \bibinfo{author}{\bibfnamefont{A.~P.} \bibnamefont{Schnyder}},
  \bibinfo{journal}{New J. Phys.} \textbf{\bibinfo{volume}{21}},
  \bibinfo{pages}{073003} (\bibinfo{year}{2019}).

\bibitem[{\citenamefont{Mullen et~al.}(2015)\citenamefont{Mullen, Uchoa, and
  Glatzhofer}}]{Mullen:2015}
\bibinfo{author}{\bibfnamefont{K.}~\bibnamefont{Mullen}},
  \bibinfo{author}{\bibfnamefont{B.}~\bibnamefont{Uchoa}}, \bibnamefont{and}
  \bibinfo{author}{\bibfnamefont{D.~T.} \bibnamefont{Glatzhofer}},
  \bibinfo{journal}{Phys. Rev. Lett.} \textbf{\bibinfo{volume}{115}},
  \bibinfo{pages}{026403} (\bibinfo{year}{2015}).

\bibitem[{\citenamefont{Rhim and Kim}(2015)}]{Rhim:2015}
\bibinfo{author}{\bibfnamefont{J.-W.} \bibnamefont{Rhim}} \bibnamefont{and}
  \bibinfo{author}{\bibfnamefont{Y.~B.} \bibnamefont{Kim}},
  \bibinfo{journal}{Phys. Rev. B} \textbf{\bibinfo{volume}{92}},
  \bibinfo{pages}{045126} (\bibinfo{year}{2015}).

\bibitem[{\citenamefont{Lim and Moessner}(2017)}]{Lim:2017}
\bibinfo{author}{\bibfnamefont{L.-K.} \bibnamefont{Lim}} \bibnamefont{and}
  \bibinfo{author}{\bibfnamefont{R.}~\bibnamefont{Moessner}},
  \bibinfo{journal}{Phys. Rev. Lett.} \textbf{\bibinfo{volume}{118}},
  \bibinfo{pages}{016401} (\bibinfo{year}{2017}).

\bibitem[{\citenamefont{Kopnin et~al.}(2011)\citenamefont{Kopnin, Heikkil\"{a},
  and Volovik}}]{Kopnin:2011}
\bibinfo{author}{\bibfnamefont{N.~B.} \bibnamefont{Kopnin}},
  \bibinfo{author}{\bibfnamefont{T.~T.} \bibnamefont{Heikkil\"{a}}},
  \bibnamefont{and} \bibinfo{author}{\bibfnamefont{G.~E.}
  \bibnamefont{Volovik}}, \bibinfo{journal}{Phys. Rev. B}
  \textbf{\bibinfo{volume}{83}}, \bibinfo{pages}{220503(R)}
  (\bibinfo{year}{2011}).

\bibitem[{\citenamefont{Li et~al.}(2016)\citenamefont{Li, Ma, Cheng, Wang, Li,
  Zhang, Li, and Chen}}]{RonghanLi:2016}
\bibinfo{author}{\bibfnamefont{R.}~\bibnamefont{Li}},
  \bibinfo{author}{\bibfnamefont{H.}~\bibnamefont{Ma}},
  \bibinfo{author}{\bibfnamefont{X.}~\bibnamefont{Cheng}},
  \bibinfo{author}{\bibfnamefont{S.}~\bibnamefont{Wang}},
  \bibinfo{author}{\bibfnamefont{D.}~\bibnamefont{Li}},
  \bibinfo{author}{\bibfnamefont{Z.}~\bibnamefont{Zhang}},
  \bibinfo{author}{\bibfnamefont{Y.}~\bibnamefont{Li}}, \bibnamefont{and}
  \bibinfo{author}{\bibfnamefont{X.-Q.} \bibnamefont{Chen}},
  \bibinfo{journal}{Phys. Rev. Lett.} \textbf{\bibinfo{volume}{117}},
  \bibinfo{pages}{096401} (\bibinfo{year}{2016}).

\bibitem[{\citenamefont{Chan et~al.}(2016)\citenamefont{Chan, Chiu, Chou, and
  Schnyder}}]{Chan:2016}
\bibinfo{author}{\bibfnamefont{Y.-H.} \bibnamefont{Chan}},
  \bibinfo{author}{\bibfnamefont{C.-K.} \bibnamefont{Chiu}},
  \bibinfo{author}{\bibfnamefont{M.~Y.} \bibnamefont{Chou}}, \bibnamefont{and}
  \bibinfo{author}{\bibfnamefont{A.~P.} \bibnamefont{Schnyder}},
  \bibinfo{journal}{Phys. Rev. B} \textbf{\bibinfo{volume}{93}},
  \bibinfo{pages}{205132} (\bibinfo{year}{2016}).

\bibitem[{\citenamefont{Wang and Nandkishore}(2017)}]{Wang:2017}
\bibinfo{author}{\bibfnamefont{Y.}~\bibnamefont{Wang}} \bibnamefont{and}
  \bibinfo{author}{\bibfnamefont{R.~M.} \bibnamefont{Nandkishore}},
  \bibinfo{journal}{Phys. Rev. B} \textbf{\bibinfo{volume}{95}},
  \bibinfo{pages}{060506(R)} (\bibinfo{year}{2017}).

\bibitem[{\citenamefont{Liu and Balents}(2017)}]{JianpengLiu:2017}
\bibinfo{author}{\bibfnamefont{J.}~\bibnamefont{Liu}} \bibnamefont{and}
  \bibinfo{author}{\bibfnamefont{L.}~\bibnamefont{Balents}},
  \bibinfo{journal}{Phys. Rev. B} \textbf{\bibinfo{volume}{95}},
  \bibinfo{pages}{075426} (\bibinfo{year}{2017}).

\bibitem[{\citenamefont{Burkov et~al.}(2011)\citenamefont{Burkov, Hook, and
  Balents}}]{Burkov:2011}
\bibinfo{author}{\bibfnamefont{A.~A.} \bibnamefont{Burkov}},
  \bibinfo{author}{\bibfnamefont{M.~D.} \bibnamefont{Hook}}, \bibnamefont{and}
  \bibinfo{author}{\bibfnamefont{L.}~\bibnamefont{Balents}},
  \bibinfo{journal}{Phys. Rev. B} \textbf{\bibinfo{volume}{84}},
  \bibinfo{pages}{235126} (\bibinfo{year}{2011}).

\bibitem[{\citenamefont{Yu et~al.}(2017)\citenamefont{Yu, Fang, Dai, and
  Weng}}]{Yu:2017}
\bibinfo{author}{\bibfnamefont{R.}~\bibnamefont{Yu}},
  \bibinfo{author}{\bibfnamefont{Z.}~\bibnamefont{Fang}},
  \bibinfo{author}{\bibfnamefont{X.}~\bibnamefont{Dai}}, \bibnamefont{and}
  \bibinfo{author}{\bibfnamefont{H.}~\bibnamefont{Weng}},
  \bibinfo{journal}{Front. Phys.} \textbf{\bibinfo{volume}{12(3)}},
  \bibinfo{pages}{127202} (\bibinfo{year}{2017}).

\bibitem[{\citenamefont{Zhou et~al.}(2019)\citenamefont{Zhou, Zhang, Zhang, Ma,
  Feng, Mokrousov, and Yao}}]{Zhou:2019}
\bibinfo{author}{\bibfnamefont{X.}~\bibnamefont{Zhou}},
  \bibinfo{author}{\bibfnamefont{R.-W.} \bibnamefont{Zhang}},
  \bibinfo{author}{\bibfnamefont{Z.}~\bibnamefont{Zhang}},
  \bibinfo{author}{\bibfnamefont{D.-S.} \bibnamefont{Ma}},
  \bibinfo{author}{\bibfnamefont{W.}~\bibnamefont{Feng}},
  \bibinfo{author}{\bibfnamefont{Y.}~\bibnamefont{Mokrousov}},
  \bibnamefont{and} \bibinfo{author}{\bibfnamefont{Y.}~\bibnamefont{Yao}},
  \bibinfo{journal}{arXiv:1903.11025}  (\bibinfo{year}{2019}).

\bibitem[{\citenamefont{Lu et~al.}(2017)\citenamefont{Lu, Luo, Li, Yang, Cao,
  Gong, and Xiang}}]{Lu:2017}
\bibinfo{author}{\bibfnamefont{J.~L.} \bibnamefont{Lu}},
  \bibinfo{author}{\bibfnamefont{W.}~\bibnamefont{Luo}},
  \bibinfo{author}{\bibfnamefont{X.~Y.} \bibnamefont{Li}},
  \bibinfo{author}{\bibfnamefont{S.~Q.} \bibnamefont{Yang}},
  \bibinfo{author}{\bibfnamefont{J.~X.} \bibnamefont{Cao}},
  \bibinfo{author}{\bibfnamefont{X.~G.} \bibnamefont{Gong}}, \bibnamefont{and}
  \bibinfo{author}{\bibfnamefont{H.~J.} \bibnamefont{Xiang}},
  \bibinfo{journal}{Chin. Phys. Lett.} \textbf{\bibinfo{volume}{34}},
  \bibinfo{pages}{057302} (\bibinfo{year}{2017}).

\bibitem[{\citenamefont{Duca et~al.}(2015)\citenamefont{Duca, Li, Reitter,
  Bloch, Schleier-Smith, and Schneider}}]{Duca:2015}
\bibinfo{author}{\bibfnamefont{L.}~\bibnamefont{Duca}},
  \bibinfo{author}{\bibfnamefont{T.}~\bibnamefont{Li}},
  \bibinfo{author}{\bibfnamefont{M.}~\bibnamefont{Reitter}},
  \bibinfo{author}{\bibfnamefont{I.}~\bibnamefont{Bloch}},
  \bibinfo{author}{\bibfnamefont{M.}~\bibnamefont{Schleier-Smith}},
  \bibnamefont{and}
  \bibinfo{author}{\bibfnamefont{U.}~\bibnamefont{Schneider}},
  \bibinfo{journal}{Science} pp. \bibinfo{pages}{288--292}
  (\bibinfo{year}{2015}).

\bibitem[{\citenamefont{Price and Cooper}(2012)}]{Price:2012}
\bibinfo{author}{\bibfnamefont{H.~M.} \bibnamefont{Price}} \bibnamefont{and}
  \bibinfo{author}{\bibfnamefont{N.~R.} \bibnamefont{Cooper}},
  \bibinfo{journal}{Phys. Rev. A} \textbf{\bibinfo{volume}{85}},
  \bibinfo{pages}{033620} (\bibinfo{year}{2012}).

\bibitem[{\citenamefont{Jotzu et~al.}(2014)\citenamefont{Jotzu, Messer,
  Desbuquois, Lebrat, Uehlinger, Greif, and Esslinger}}]{Jotzu:2014}
\bibinfo{author}{\bibfnamefont{G.}~\bibnamefont{Jotzu}},
  \bibinfo{author}{\bibfnamefont{M.}~\bibnamefont{Messer}},
  \bibinfo{author}{\bibfnamefont{R.}~\bibnamefont{Desbuquois}},
  \bibinfo{author}{\bibfnamefont{M.}~\bibnamefont{Lebrat}},
  \bibinfo{author}{\bibfnamefont{T.}~\bibnamefont{Uehlinger}},
  \bibinfo{author}{\bibfnamefont{D.}~\bibnamefont{Greif}}, \bibnamefont{and}
  \bibinfo{author}{\bibfnamefont{T.}~\bibnamefont{Esslinger}},
  \bibinfo{journal}{Nature} pp. \bibinfo{pages}{237--240}
  (\bibinfo{year}{2014}).

\bibitem[{\citenamefont{Wang et~al.}(2006)\citenamefont{Wang, Yates, Souza, and
  Vanderbilt}}]{Wang:2006}
\bibinfo{author}{\bibfnamefont{X.}~\bibnamefont{Wang}},
  \bibinfo{author}{\bibfnamefont{J.~R.} \bibnamefont{Yates}},
  \bibinfo{author}{\bibfnamefont{I.}~\bibnamefont{Souza}}, \bibnamefont{and}
  \bibinfo{author}{\bibfnamefont{D.}~\bibnamefont{Vanderbilt}},
  \bibinfo{journal}{Phys. Rev. B} p. \bibinfo{pages}{195118}
  (\bibinfo{year}{2006}).

\bibitem[{\citenamefont{Marzari et~al.}(2012)\citenamefont{Marzari, Mosto,
  Yates, Souza, and Vanderbilt}}]{Marzari:2012}
\bibinfo{author}{\bibfnamefont{N.}~\bibnamefont{Marzari}},
  \bibinfo{author}{\bibfnamefont{A.~A.} \bibnamefont{Mosto}},
  \bibinfo{author}{\bibfnamefont{J.~R.} \bibnamefont{Yates}},
  \bibinfo{author}{\bibfnamefont{I.}~\bibnamefont{Souza}}, \bibnamefont{and}
  \bibinfo{author}{\bibfnamefont{S.}~\bibnamefont{Vanderbilt}},
  \bibinfo{journal}{Rev. Mod. Phys.} pp. \bibinfo{pages}{1419--1475}
  (\bibinfo{year}{2012}).

\bibitem[{\citenamefont{Gradhand et~al.}(2012)\citenamefont{Gradhand, Fedorov,
  Pientka, Zahn, Mertig, and Gy\"orffy}}]{Gradhand:2012}
\bibinfo{author}{\bibfnamefont{M.}~\bibnamefont{Gradhand}},
  \bibinfo{author}{\bibfnamefont{D.~V.} \bibnamefont{Fedorov}},
  \bibinfo{author}{\bibfnamefont{F.}~\bibnamefont{Pientka}},
  \bibinfo{author}{\bibfnamefont{P.}~\bibnamefont{Zahn}},
  \bibinfo{author}{\bibfnamefont{I.}~\bibnamefont{Mertig}}, \bibnamefont{and}
  \bibinfo{author}{\bibfnamefont{B.~L.} \bibnamefont{Gy\"orffy}},
  \bibinfo{journal}{J. Phys. Condens. Matter} p. \bibinfo{pages}{213202}
  (\bibinfo{year}{2012}).

\bibitem[{\citenamefont{Thonhauser et~al.}(2005)\citenamefont{Thonhauser,
  Ceresoli, Vanderbilt, and Resta.}}]{Thonhauser:2005}
\bibinfo{author}{\bibfnamefont{T.}~\bibnamefont{Thonhauser}},
  \bibinfo{author}{\bibfnamefont{D.}~\bibnamefont{Ceresoli}},
  \bibinfo{author}{\bibfnamefont{D.}~\bibnamefont{Vanderbilt}},
  \bibnamefont{and} \bibinfo{author}{\bibfnamefont{R.}~\bibnamefont{Resta.}},
  \bibinfo{journal}{Phys. Rev. Lett.} p. \bibinfo{pages}{137205}
  (\bibinfo{year}{2005}).

\bibitem[{\citenamefont{Xiao et~al.}(2005)\citenamefont{Xiao, Shi, and
  Niu}}]{Xiao:2005}
\bibinfo{author}{\bibfnamefont{D.}~\bibnamefont{Xiao}},
  \bibinfo{author}{\bibfnamefont{J.}~\bibnamefont{Shi}}, \bibnamefont{and}
  \bibinfo{author}{\bibfnamefont{Q.}~\bibnamefont{Niu}},
  \bibinfo{journal}{Phys. Rev. Lett.} p. \bibinfo{pages}{137204}
  (\bibinfo{year}{2005}).

\bibitem[{\citenamefont{Ceresoli et~al.}(2006)\citenamefont{Ceresoli,
  Thonhauser, Vanderbilt, and Resta}}]{Ceresoli:2006}
\bibinfo{author}{\bibfnamefont{D.}~\bibnamefont{Ceresoli}},
  \bibinfo{author}{\bibfnamefont{T.}~\bibnamefont{Thonhauser}},
  \bibinfo{author}{\bibfnamefont{D.}~\bibnamefont{Vanderbilt}},
  \bibnamefont{and} \bibinfo{author}{\bibfnamefont{R.}~\bibnamefont{Resta}},
  \bibinfo{journal}{Phys. Rev. B} p. \bibinfo{pages}{024408}
  (\bibinfo{year}{2006}).

\bibitem[{\citenamefont{Shi et~al.}(2007)\citenamefont{Shi, Vignale, Xiao, and
  Niu}}]{Shi:2007}
\bibinfo{author}{\bibfnamefont{J.}~\bibnamefont{Shi}},
  \bibinfo{author}{\bibfnamefont{G.}~\bibnamefont{Vignale}},
  \bibinfo{author}{\bibfnamefont{D.}~\bibnamefont{Xiao}}, \bibnamefont{and}
  \bibinfo{author}{\bibfnamefont{Q.}~\bibnamefont{Niu}},
  \bibinfo{journal}{Phys. Rev. Lett.} p. \bibinfo{pages}{197202}
  (\bibinfo{year}{2007}).

\bibitem[{\citenamefont{Souza and Vanderbilt}(2008)}]{Souza:2008}
\bibinfo{author}{\bibfnamefont{I.}~\bibnamefont{Souza}} \bibnamefont{and}
  \bibinfo{author}{\bibfnamefont{D.}~\bibnamefont{Vanderbilt}},
  \bibinfo{journal}{Phys. Rev. B} p. \bibinfo{pages}{054438}
  (\bibinfo{year}{2008}).

\bibitem[{\citenamefont{Dittrich et~al.}(1998)\citenamefont{Dittrich,
  H{\"a}nggi, Ingold, Kramer, Sch{\"o}n, and Zwerger}}]{Haenggi}
\bibinfo{author}{\bibfnamefont{T.}~\bibnamefont{Dittrich}},
  \bibinfo{author}{\bibfnamefont{P.}~\bibnamefont{H{\"a}nggi}},
  \bibinfo{author}{\bibfnamefont{G.-L.} \bibnamefont{Ingold}},
  \bibinfo{author}{\bibfnamefont{B.}~\bibnamefont{Kramer}},
  \bibinfo{author}{\bibfnamefont{G.}~\bibnamefont{Sch{\"o}n}},
  \bibnamefont{and} \bibinfo{author}{\bibfnamefont{W.}~\bibnamefont{Zwerger}},
  \emph{\bibinfo{title}{Quantum Transport and Dissipation}}
  (\bibinfo{publisher}{Wiley-VCH}, \bibinfo{year}{1998}).

\bibitem[{\citenamefont{Bott and Seeley}(1978)}]{Bott:1978}
\bibinfo{author}{\bibfnamefont{R.}~\bibnamefont{Bott}} \bibnamefont{and}
  \bibinfo{author}{\bibfnamefont{R.}~\bibnamefont{Seeley}},
  \bibinfo{journal}{Commun. Math. Phys.} \textbf{\bibinfo{volume}{62}},
  \bibinfo{pages}{235} (\bibinfo{year}{1978}).

\end{thebibliography}

\end{document}